\documentclass[sigconf,natbib=true]{acmart}
\usepackage{multirow}
\usepackage{amsfonts}
\usepackage{float}
\usepackage{url}
\usepackage{amsmath}
\usepackage{algorithm}
\usepackage{algpseudocode}
\usepackage{bm}
\usepackage{geometry}
\usepackage{graphicx}
\usepackage{caption}
\usepackage[skip=0cm,list=true,labelfont=it]{subcaption}
\usepackage{xcolor}

\usepackage{booktabs}
\usepackage{enumitem} 
\usepackage{ccicons}

\theoremstyle{definition}

\AtBeginDocument{%
  \providecommand\BibTeX{{%
    \normalfont B\kern-0.5em{\scshape i\kern-0.25em b}\kern-0.8em\TeX}}}

\copyrightyear{2025}
\acmYear{2025}
\setcopyright{cc}
\setcctype{by}
\acmConference[RecSys '25]{Proceedings of the Nineteenth ACM Conference on Recommender Systems}{September 22--26, 2025}{Prague, Czech Republic}
\acmBooktitle{Proceedings of the Nineteenth ACM Conference on Recommender Systems (RecSys '25), September 22--26, 2025, Prague, Czech Republic}
\acmDOI{10.1145/3705328.3748167}
\acmISBN{979-8-4007-1364-4/2025/09}

\begin{document}

\title{Exploring the Potential of LLMs for Serendipity Evaluation in Recommender Systems}

\author{Li Kang}
\affiliation{
  \institution{Hong Kong Baptist University}
  \city{Hong Kong}
  \country{China}
}
\email{likang@comp.hkbu.edu.hk}
\authornote{Equal contributions.}

\author{Yuhan Zhao}
\affiliation{
  \institution{Hong Kong Baptist University}
  \city{Hong Kong}
  \country{China}
}
\email{csyhzhao@comp.hkbu.edu.hk}
\authornotemark[1] 

\author{Li Chen}
\affiliation{
  \institution{Hong Kong Baptist University}
  \city{Hong Kong}
  \country{China}
}
\email{lichen@comp.hkbu.edu.hk}

\begin{abstract}
Serendipity plays a pivotal role in enhancing user satisfaction within recommender systems, yet its evaluation poses significant challenges due to its inherently subjective nature and conceptual ambiguity. Current algorithmic approaches predominantly rely on proxy metrics for indirect assessment, often failing to align with real user perceptions, thus creating a gap. With large language models (LLMs) increasingly revolutionizing evaluation methodologies across various human annotation tasks, we are inspired to explore a core research proposition: \textit{Can LLMs effectively simulate human users for serendipity evaluation?}

To address this question, we conduct a meta-evaluation on two datasets derived from real user studies in the e-commerce and movie domains, focusing on three key aspects: the accuracy of LLMs compared to conventional proxy metrics, the influence of auxiliary data on LLM comprehension, and the efficacy of recently popular multi-LLM techniques. Our findings indicate that even the simplest zero-shot LLMs achieve parity with, or surpass, the performance of conventional metrics. Furthermore, multi-LLM techniques and the incorporation of auxiliary data further enhance alignment with human perspectives. Based on our findings, the optimal evaluation by LLMs yields a Pearson correlation coefficient of 21.5\% when compared to the results of the user study. 
This research implies that LLMs may serve as potentially accurate and cost-effective evaluators, introducing a new paradigm for serendipity evaluation in recommender systems. Our code is publicly available at \url{https://github.com/Leah-HKBU/SerenEva}.
\end{abstract}

\maketitle
\section{Introduction}
Serendipity plays a pivotal role in enhancing user satisfaction within recommender systems by mitigating the effects of the information cocoon and filter bubble issues from traditional recommendation methods~\cite{SerenCDR, SerenPrompt, PURS, hasan2023topic, wang2024llms}. 
Despite growing recognition of serendipity's importance and subsequent algorithmic innovations, how to evaluate serendipity remains challenging. This difficulty arises from its inherently subjective nature and conceptual ambiguity, distinguishing it from traditional accuracy-oriented metrics~\cite{fu2023deep, ziarani2021serendipity, taobao}. 
The gold standard for user-centered evaluation involves carefully designed user studies that directly capture user feedback~\cite{kotkov2023rethinking, Binst, pu2011user}, which, however, are costly in practice. As a result, many researchers rely on predefined proxy metrics (e.g., relevance-unexpectedness) to approximate serendipity scores, which are then treated as ground truth for algorithm evaluation~\cite{DESR, PURS, SNPR}. 
Nonetheless, the gap between these indirect measurements and actual user perceptions introduces bias into serendipity research~\cite{kotkov2024dark, taobao}.

The emergence of large language models (LLMs) has revolutionized evaluation methodologies across human annotation tasks, showcasing remarkable potential in user simulation and automatic assessment~\cite{bavaresco2024llms, ashktorab2024aligning, tseng2024expert, li2024llms}. This breakthrough motivates our key research question: \textbf{\textit{Can LLMs effectively simulate human users for serendipity evaluation?}} 
If the answer is yes, the LLM-based evaluation approach could effectively combine the accuracy of user studies with the efficiency of proxy metrics, potentially transforming research paradigms in serendipity.
While preliminary explorations have been conducted~\cite{Japan, SerenPrompt}, to our knowledge, no comprehensive studies have addressed the effectiveness of various LLM techniques across different product domains and data types for serendipity evaluation. 
To systematically investigate this proposition, we have attempted to address the following three research questions:  
\begin{itemize}
    \item \textit{RQ1: Can LLMs using basic prompt strategies surpass conventional proxy metrics in serendipity evaluation?}
    \item \textit{RQ2: What auxiliary data (e.g., user age and gender) might further enhance the LLMs' understanding of serendipity?}
    \item \textit{RQ3: Can advanced multi-LLM techniques help improve the accuracy of evaluation?}
\end{itemize}

To address these questions, we employ two user-study-validated serendipity datasets from distinct domains: e-commerce and movies. This allows us to identify both general and domain-specific observations, enhancing the comprehensiveness and credibility of our conclusions. Then, we propose SerenEva (\textbf{seren}dipity \textbf{eva}luation framework), a novel meta-evaluation framework that measures the alignment between LLM evaluator ratings and human judgments using correlation and error metrics.

For RQ1, we benchmark zero-shot and few-shot LLM evaluators (e.g., Qwen2.5-7B~\cite{yang2024qwen2}) against conventional proxy metrics using basic prompting strategies. To address RQ2, we categorize and inject auxiliary data through structured prompting. For RQ3, we explore the effectiveness of multi-LLM techniques. 
Our findings reveal: 
\begin{itemize}
    \item Zero-shot and few-shot LLMs achieve parity with or surpass conventional metrics across various evaluation dimensions (e.g., Pearson correlation coefficient), demonstrating their viable potential as serendipity evaluators. 
    \item The incorporation of auxiliary data (e.g., user curiosity and item similarity) markedly enhances the accuracy of LLM evaluations, although the optimal choice of auxiliary data is contingent upon the specific domain. 
    \item Multi-LLM techniques with a score averaging strategy substantially improve evaluation performance and alignment with human judgments.
\end{itemize}

Based on these findings, the optimal evaluation by LLMs yields a Pearson correlation coefficient of over 20\% when compared to the results of the user study, implying the potential of using LLMs to evaluate recommendation serendipity.

In summary, our contributions include investigating the feasibility of leveraging LLMs as evaluators for serendipity in recommender systems and exploring the incorporation of auxiliary data and multi-LLM techniques to enhance their understanding of serendipity. We demonstrate that LLMs, when appropriately prompted, may serve as potentially reliable and reproducible evaluators that surpass conventional proxy metrics in accuracy. 

\section{Experimental Setup}

\subsection{Problem Formulation} 
Let $\mathcal{U}$ ($|\mathcal{U}| = M$) and $\mathcal{V}$ ($|\mathcal{V}| = N$) denote the sets of users and items, respectively. Traditional recommender systems focus on recommending a subset of items $\mathcal{M} \subseteq \mathcal{V}$ to maximize accuracy metrics such as NDCG and Recall~\cite{ZCH24, ZCC25}. In contrast, serendipity-oriented recommendations introduce an additional criterion: the ability to recommend unexpected yet relevant items~\cite{taobao}. Our research problem centers on identifying which method might approximate the ground truth obtained from user studies.

\subsection{User Study as Gold Standard}
We obtained two public serendipity datasets validated through rigorous user studies, with dataset statistics shown in Table~\ref{tab:statistics}:
\begin{itemize}
    \item \textit{Taobao Serendipity}~\cite{taobao}, which was collected through a user survey on Mobile Taobao, a leading e-commerce platform in China. It captures user perceptions of serendipity ("pleasant surprise") on 5-point Likert scales, along with users' demographic profiles (e.g., age and gender) and psychological profiles including curiosity (CEI-II~\cite{kashdan2009curiosity}) and Big-Five personality traits (TIPI~\cite{gosling2003very}).
    \item \textit{Serendipity-2018}~\cite{seren2018}, which was collected through a user survey on the MovieLens platform. The survey captures users' perceptions of serendipity through eight statements rated on 5-point Likert scales, covering aspects such as movie discovery, novelty, unexpectedness, and preference broadening. Although the dataset lacks direct user ratings on serendipity, we calculated the score by averaging three unexpectedness-related variables, as validated in~\cite{wang2023item}.
\end{itemize}
By examining two distinct product domains, e-commerce and movies, we aim to identify general and domain-specific insights, thereby enhancing the comprehensiveness of our conclusions. Because user studies provide real user feedback, making them the closest approximation to actual user perceptions. 
Therefore, we regard these results as the gold standard for evaluating various simulators (also called evaluators) in this work. 

Another commonly used dataset is SerenLens~\cite{fu2023wisdom}, which includes annotations from third-party summarization. We do not use this dataset because our work focuses on using LLMs as user simulators, and user studies give us direct feedback from end users, which better matches our goals. 

\begin{table}
\centering
\caption{Statistics of two serendipity datasets.}
\label{tab:statistics}
\begin{tabular}{l l r r r}
\toprule
\textbf{Dataset} & \textbf{Domain} & \textbf{\#Users} & \textbf{\#Items} & \textbf{\#Ratings} \\
\midrule
Taobao Serendipity & Shopping & 11,383 & 9,985 & 11,383 \\
Serendipity-2018 & Movie & 481 & 1,678 & 2,150 \\
\bottomrule
\end{tabular}
\begin{flushleft}
\footnotesize{\textbf{Note:} The numbers of ratings refer specifically to serendipity ratings.}
\end{flushleft}
\end{table}

\subsection{Proxy Metrics for Serendipity}
Due to the cost and time constraints associated with user studies, many researchers have adopted proxy metrics to indirectly evaluate serendipity~\cite{PURS,DESR,SNPR,kotkov2020does}. These metrics do not incorporate real users' attitudes, but instead calculate serendipity scores based on their assumptions. With those scores, the evaluation transforms into a ranking task~\cite{PURS, kotkov2020does} or regression task~\cite{SNPR}, using metrics such as NDCG or HR for the final assessment. 
For convenience, we use acronyms from the original papers to refer to these metrics.

\paragraph{SOG (Serendipity-Oriented Greedy)~\cite{kotkov2020does}}
This method integrates multiple dimensions:
\begin{equation}
S_{uiB} = \sum_{\phi \in \Phi} \alpha_\phi \cdot \phi(uiB)
\end{equation}
where $\Phi = \{\text{relevance}, \text{diversity}, \text{history dissimilarity}, \text{unpopularity}\}$ with corresponding weights $\alpha_\phi$.

\paragraph{SNPR (Serendipity-oriented Next POI Recommendation)~\cite{SNPR}}
In this work, serendipity is defined as a linear combination of relevance and unexpectedness:
\begin{equation}
\text{Serendipity}(i,u) = \lambda\text{R}(i,u) + (1-\lambda)\text{U}(i,u)
\end{equation}
where $R(i,u)$ represents relevance, which can be inferred from user interactions, and $U(i,u)$ represents unexpectedness, calculated using multi-level dissimilarity measures.

\paragraph{PURS (Personalized Unexpected Recommender System)~\cite{PURS}}
A hybrid utility function is proposed in this work to integrate the unexpectedness factor:
\begin{equation}
\text{Utility}_{u,i} = r_{u,i} + f(\text{unexp}_{u,i}) \cdot \text{unexp\_factor}_{u,i}
\end{equation}
where unexpectedness derives from the embedding distance to the user interest clusters.


\paragraph{DESR (Directional and Explainable Serendipity Recommendation)~\cite{DESR}}
It adopts an F-score-inspired formulation:
\begin{equation}
\text{AD} = \frac{\text{acc} \cdot \text{dif}}{\text{acc} + \text{dif}}
\end{equation}
where accuracy ($\text{acc}$) combines long- and short-term preference alignment, and $\text{dif}$ balances diversity and historical dissimilarity.

\subsection{LLM-Based Evaluation Framework}
Our LLM evaluator employs constrained prompting strategies to assess the serendipity of recommendation items. The basic \textbf{prompt} is shown below. 


\noindent
\setlength{\fboxsep}{4pt} 
\setlength{\fboxrule}{1.1pt} 
\begin{center} 
\colorbox{gray!5!white}{
\textcolor{gray!70!black}{
    \fbox{
        \begin{minipage}{0.90\columnwidth}
            \vspace{2pt} 
            \textcolor{black}{
You are a user of a Chinese e-commerce platform, and you have received a user survey that aims to gather your opinion on the serendipity of the items recommended to you. Serendipity here means that the item recommended is a pleasant surprise.\\
\#\# Background\\
You have used the Chinese e-commerce platform, and this recommendation is based on your behavior history. You are provided with the genres of the recommended item and the items you have clicked on or purchased. Your behavior history is listed in a comma-separated format, sorted from oldest to newest.\\
\#\# Task\\
Please provide a serendipity rating for the recommended item using a 5-point Likert scale: 1 – ``strongly disagree''; 2 – ``disagree''; 3 – ``neither agree nor disagree''; 4 – ``agree''; 5 – ``strongly agree''. \\
Rate the recommended item from the perspective of serendipity, based on your behavior history.\\
\#\# Output Format\\
Generate only the rating number, without any additional commentary or explanation.\\
\#\# Response\\
Your behavior history: [\textit{user behavior history}] \\
Recommended item: (\textit{item info}) \\
Your serendipity rating:}
            \vspace{2pt} 
        \end{minipage}%
        }
    }%
}
\end{center}

Additional details and the remainder of the prompt are available in our open-source codebase. All experiments were conducted using publicly available models (e.g., LLaMA2~\cite{llama2}, Qwen2.5~\cite{yang2024qwen2}, GPT-4~\cite{achiam2023gpt}) to ensure transparency and reproducibility. 

\subsection{Meta-Evaluation Protocol (SerenEva)}
One of the most widely used methods for assessing the efficacy of an evaluator is meta-evaluation~
\cite{zhang2024large}. 
To facilitate the meta-evaluation of serendipity in our case, we propose SerenEva, which includes an evaluation function $h(\cdot)$:
\begin{align}
    h = \frac{1}{|\mathcal{U}|}\sum_{u \in \mathcal{U}} r(s_{*},s_{real})
\end{align}
where $s_{real}$ denotes the real user feedback. $s_*$ represents serendipity scores derived from different evaluation methods (e.g., $s_{\text{LLM}}$ stands for evaluation results based on LLMs). $h$ measures the discrepancy between a particular evaluation method's evaluation result $s_{*}$ and the user feedback $s_{real}$ across various indicators $r$. To ensure a rigorous and comprehensive comparison, we introduce three major metrics: Pearson correlation coefficient, mean absolute error (MAE), and root mean squared error (RMSE). The Pearson correlation coefficient can capture linear relationships. MAE provides robustness against outliers, and RMSE imposes stricter penalties for larger deviations. If RMSE is much larger than MAE, it indicates that the LLM performs well for the majority of users but exhibits large errors for some users. 

SerenEva requires the unification of input format and scope to align with the user feedback format. While LLM predictions can be obtained directly through constrained prompts, proxy metrics require certain adjustments. Since the ground truth in both datasets is based on a 5-point Likert scale (1 -- ``strongly disagree''; 2 -- ``disagree''; 3 -- ``neither agree nor disagree''; 4 -- ``agree''; 5 -- ``strongly agree''), we apply the following transformation to achieve this format:

\begin{equation}
\text{score} = \text{round}\left(\frac{\text{output} - \text{min\_output}}{\text{max\_output} - \text{min\_output}} \times 4 + 1\right)
\end{equation}
$\text{max\_output}$ and $\text{min\_output}$ represent the maximum and minimum values of the metric's predictions across all outputs, respectively. $\text{output}$ is the raw prediction value from the metric, and $\text{round}(\cdot)$ denotes rounding to the nearest integer. 
The final $\text{score}$ is thus mapped to the 5-point Likert scale, ensuring compatibility with the ground truth rating.

Due to space constraints, we present only the experimental results that demonstrate statistically significant improvements over the strongest baseline(s), as determined by a two-sided t-test with $p < 0.05$. Additionally, all reported Pearson correlation coefficients are statistically significant at the $p < 0.05$ level. To ensure the stability and reproducibility of our findings, we used a low temperature setting (0.00001) and averaged the results over five runs.

\begin{table*}
\centering
\caption{Performance comparison of conventional proxy metrics and LLMs. 
The best results are highlighted in \textbf{bold}, and the second-best results are \underline{underlined}. SerenPrompt\_1 and SerenPrompt\_2 represent Discrete Style 1 and 2 from~\cite{SerenPrompt}, respectively.}
\label{tab:RQ1 experiment_results}
{
\begin{tabular}{l|ccc|ccc}
\toprule
\multirow{2}{*}{Method} & \multicolumn{3}{c|}{Taobao Serendipity} & \multicolumn{3}{c}{Serendipity-2018} \\
\cmidrule(lr){2-4} \cmidrule(lr){5-7}
                         & Pearson(\%) & MAE & RMSE & Pearson(\%) & MAE & RMSE \\
\midrule
Random & -0.5771 & 1.6484 & 2.0637 & -0.1329 & 1.4400 & 1.7847 \\
\midrule
\multicolumn{7}{c}{\textit{Proxy Metrics}} \\ 
\midrule
SOG & 4.6095& 1.5231 & 1.8736 & 5.7226 & 1.2746 & 1.5725 \\
PURS & 2.6744 & 1.6540 & 2.0356& 3.3270 & 1.2604 & 1.5608  \\
DESR & 3.8458 & 1.6132 & 1.9675& 3.4840 & 1.1170 & 1.4059  \\
SNPR & 0.9033 & 1.5969 & 1.9955& 0.0509 & 1.1062 & 1.3824  \\
\midrule
\multicolumn{7}{c}{\textit{LLM-based Baseline}} \\ 
\midrule
LLM4Seren & 4.1338 & 1.4286 & 1.7603& -7.6891 & 1.1084 & 1.4113   \\
SerenPrompt\_1 & 2.2306 & 1.5063 & 1.8764& 6.0190 & 1.0165 & 1.3215  \\
SerenPrompt\_2 & 9.2422 & 1.4271 & 1.8103& 8.3274 & 1.0773 & 1.3936  \\
\midrule
\multicolumn{7}{c}{\textit{LLM-based inference (Zero-Shot)}} \\ 
\midrule
LLaMA2-7B & 0.2766 & 1.6311 & 1.9910& -0.0802 & 1.1464 & 1.4653  \\
LLaMA2-13B & 1.4615 & 1.6236 & 1.9809& 0.6997 & 1.5564 & 1.8794  \\
Qwen2.5-7B & 7.5001 & 1.5081 & 1.8401&  4.1820 & 1.3485 & 1.5764  \\
Qwen2.5-14B & 7.1590 & 1.4153 & 1.7926& 6.2096 & 1.2507 & 1.5655  \\
Qwen2.5-72B & 10.5165 & \textbf{1.3263} & \textbf{1.6601}& 10.3220 & 1.1885 & 1.4678  \\
GPT-4 &10.8139  &1.5316  &1.8631 & 10.8859 & 1.0126 & 1.3292  \\
\midrule
\multicolumn{7}{c}{\textit{LLM-based inference (Few-Shot)}} \\ 
\midrule
LLaMA2-7B & 1.9549 & 1.5694 & 1.9687& 3.0739 & 1.0690 & 1.3945  \\
LLaMA2-13B & 1.9945 & 1.5696 & 1.9689& 6.7423 & 1.4150 & 1.6949  \\
Qwen2.5-7B & 10.5019 & 1.3836 & 1.7749& 6.4900 & 1.0068 & \underline{1.2819}  \\
Qwen2.5-14B & 10.2701 & 1.4018 & 1.7836 &\underline{11.4769} &\underline{0.9742} & 1.2825  \\
Qwen2.5-72B &\underline{11.9836} & \underline{1.3756} & \underline{1.7734}& 8.3117 &\textbf{0.8591} &\textbf{1.1830} \\
GPT-4 & \textbf{12.1231} & 1.5111 & 1.8405 &\textbf{14.5545} & 1.0112 & 1.3436  \\
\bottomrule
\end{tabular}}
\end{table*}

\section{RQ1: LLM as a Competitive Evaluator}
\label{sec:rq1}

In this section, we explore whether LLMs can surpass conventional proxy metrics in evaluating serendipity, even in the simplest settings. 
To this end, we exclusively rely on the user's historical behavior data, focusing specifically on the last 10 items interacted with by the user prior to the target item in both datasets. 
The results are presented in Table~\ref{tab:RQ1 experiment_results}.

\subsection{Limitations of Conventional Proxy Metrics in Capturing User Feedback on Serendipity}

As expected, proxy metrics surpass random baselines, indicating their ability to partially capture user serendipity perception. 
Notably, the \textbf{SOG} metric~\cite{kotkov2020does} demonstrates the best performance among all proxy metrics, likely due to its comprehensive integration of relevance, popularity, and diversity. 
However, SOG still shows a significant gap compared to real user feedback, likely due to its use of a fixed weighting strategy for different components.

Furthermore, most metrics exhibit consistent performance, except SNPR~\cite{SNPR}. On the Serendipity-2018 dataset, despite poor Pearson correlation results, SNPR performs well in terms of MAE and RMSE.  
This inconsistency stems from SNPR's design, which heavily weights relevance (70\%) in its calculations. When recommendations are serendipitous, they may contain unexpected items that exhibit lower relevance scores. These cases tend to become outliers in SNPR's assessment, particularly affecting Pearson correlation calculations, while MAE and RMSE remain more robust as they measure absolute differences rather than linear relationships.

\subsection{Potential of LLMs as Evaluators}

In zero-shot settings, LLMs such as Qwen2.5-14B and GPT-4 surpass conventional metrics across both datasets by approximately 100\% in Pearson correlation compared to the best proxy metric (\textbf{SOG}). 
We attribute this to LLMs' robust background knowledge and user simulation capabilities, underscoring their potential as competent evaluators in the serendipity evaluation context.

In few-shot settings with five examples, most LLMs (e.g., Qwen family and GPT-4) demonstrate superior performance compared to that in zero-shot settings. 
Remarkably, even the smaller parameter model Qwen2.5-7B achieves an impressive performance score of 10.50\% on the Taobao dataset, approaching the performance of Qwen2.5-72B in zero-shot scenarios. This phenomenon inspires us to consider that, in the future, we can effectively enhance the evaluation capabilities of LLMs by incorporating limited user study data. More importantly, these results underscore the potential of smaller parameter models to deliver high-quality outcomes in few-shot learning scenarios. They offer significant advantages in terms of time and cost efficiency, making them a compelling and practical choice for various applications.

Regrettably, the LLaMA family models do not exhibit exceptional performance, especially on the Taobao dataset. This limitation may stem from the dataset's focus on the Chinese community and its inclusion of product descriptions in Chinese. Previous research has pointed out LLaMA2's limited proficiency in processing Chinese~\cite{zhao2024llama,wendler2024llamas}, which may explain its subpar performance in this scenario.

\subsection{LLM-based Methods} 
SerenPrompt~\cite{SerenPrompt} aims to leverage the inference capability of LLMs to predict the serendipity level. By utilizing its prompts, we explore the application of this method for evaluation purposes. LLM4Seren~\cite{Japan} introduces a straightforward LLM-based approach for assessing recommendation serendipity and validates the evaluation capability of LLMs and the \textbf{SOG} metric, which serves as a proxy metric we have discussed.

Our results, based on Qwen2.5-14B~\cite{yang2024qwen2}, show that SerenPrompt performs well, likely due to its accommodation of both unexpectedness and relevance in prompts, which are two crucial elements of serendipity. 
However, there is still a performance gap compared to our method. The core reason for this gap might be that SerenPrompt optimizes the recommendation task rather than being specifically designed as an evaluator, thus resulting in a natural disparity in final performance. Conversely, LLM4Seren performs poorly, possibly due to its lack of an explicit serendipity definition in prompts, leading to suboptimal evaluation capability. 
These observations drive us to carefully design and explore the use of LLMs as evaluators of recommendation serendipity.

\section{RQ2: The Role of Auxiliary Data in Enhancing LLM-based Serendipity Evaluation}
\label{sec:data}
In this section, we leverage Qwen2.5-14B~\cite{yang2024qwen2} to investigate how different types of auxiliary data can influence LLM-based serendipity evaluation performance. Our goal is to identify which type(s) of data might be useful for enhancing the model's evaluation accuracy. Throughout this section, our default configuration uses five-shot prompting with the 10 most recent user interactions prior to the target item, with ``NA'' denoting the baseline condition without auxiliary data.
Based on the work of~\cite{wang2023item, SNPR, xi2025bursting}, we categorize the auxiliary data into three distinct types: user data, item data, and interaction data, as summarized in Table \ref{tab:Auxiliary Data}. User data comprises psychological, demographic, and profile data, capturing a comprehensive view of individual characteristics and behaviors. 
Item data includes key attributes of items, such as popularity and item similarity. 
Interaction data focuses on two key dimensions: (1) the quantity of interaction data used, and (2) the types of interactions, such as clicks, purchases, and ratings.

Furthermore, we also explore the impact of category-level data~\cite{wang2023item}
that captures user preferences for broader item categories (such as weekly interaction frequency, hourly interaction frequency, time since last interaction, and category interaction frequency). However, our experiments revealed that adding such data types does not significantly improve evaluation performance. Due to space limitations, we thus omit this part of the experimental results.

\begin{table}
\centering
      \caption{Auxiliary Data}
    \label{tab:Auxiliary Data}
    \small
    \begin{tabular}{@{}p{0.3\linewidth}p{0.65\linewidth}@{}}
    \toprule
    \textbf{Data type}& \textbf{Description}\\
    \midrule
\multicolumn{2}{c}{\textit{User data}} \\
\midrule
    \textbf{Psychological Data}  & Curiosity and Big-Five personality traits\\
    \textbf{Demographic Data} & Age and Gender\\
        \textbf{Profile Data} &  Long-term profile and Short-term profile respectively containing long and recent interactions \\
    \midrule
\multicolumn{2}{c}{\textit{Item data}} \\ 
\midrule
    \textbf{Popularity}&  The item's popularity\\
    \textbf{Similarity}&  The item's similarity to others\\
    \midrule
\multicolumn{2}{c}{\textit{Interaction data}} \\ 
\midrule
    \textbf{Interaction Length} &  The length of historical interactions used\\ 
    \textbf{Interaction Type} & The type of interaction, such as clicks, ratings.\\
    \bottomrule
    \end{tabular}

\end{table}

\begin{figure}
    \centering
    \begin{subfigure}[b]{0.48\columnwidth} 
        \includegraphics[width=\textwidth]{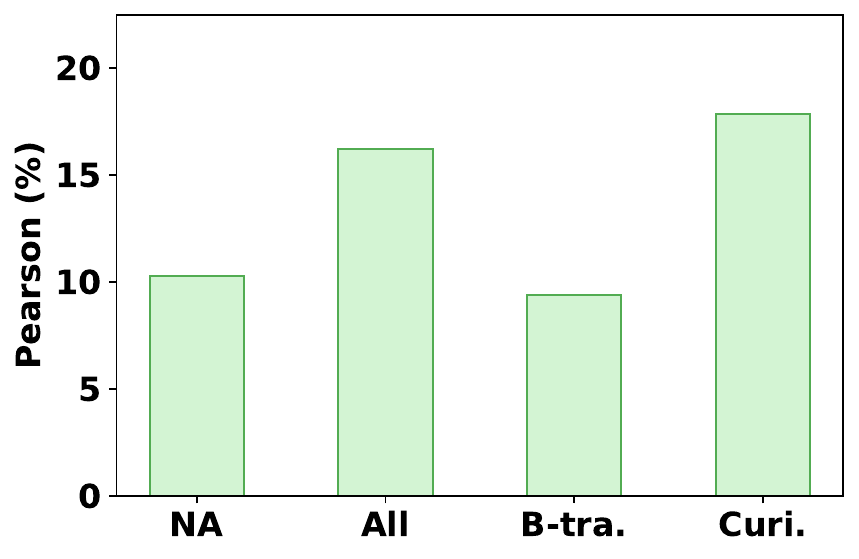}
        \caption{Psychological Data}
        \label{fig:psy}
    \end{subfigure}
    \hfill 
    \begin{subfigure}[b]{0.48\columnwidth} 
        \includegraphics[width=\textwidth]{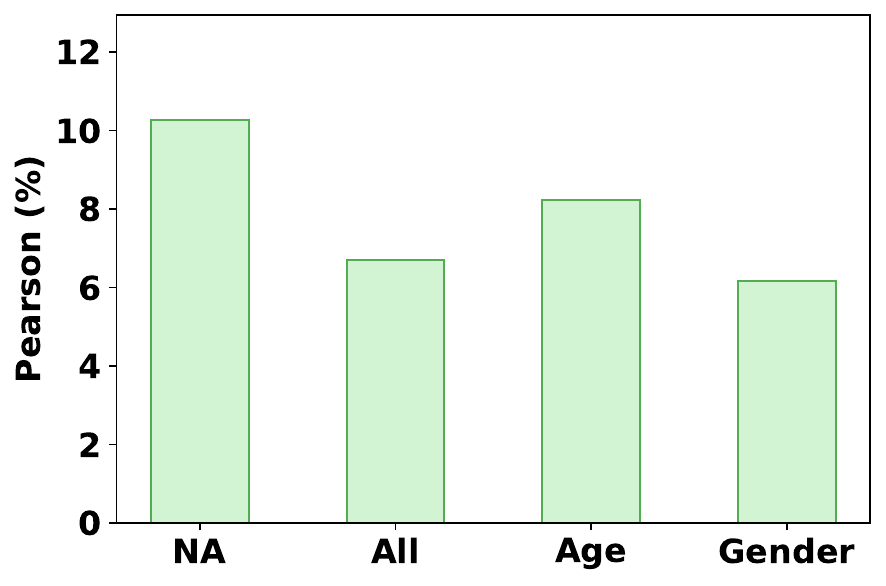}
        \caption{Demographic Data}
        \label{fig: demographic features}
    \end{subfigure}
    \caption{Comparison w.r.t. Pearson correlation coefficient for different user attributes in the Taobao dataset (B-tra.: Big Five Personality Traits, Curi.: Curiosity, All: Combined data within each figure).}
    \label{fig:comparison_demo_psy}
\end{figure}

\begin{figure*} 
    \centering
    \begin{subfigure}[b]{0.32\textwidth}
        \includegraphics[width=\textwidth]{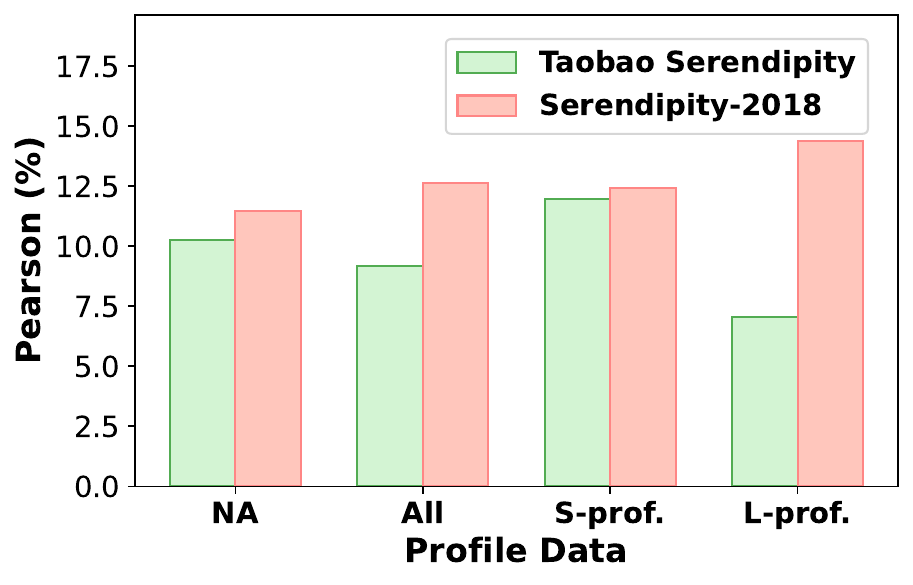}
        \caption{Pearson correlation coefficient}
        \label{fig:pearson_comparison}
    \end{subfigure}
    \hfill 
    \begin{subfigure}[b]{0.32\textwidth}
        \includegraphics[width=\textwidth]{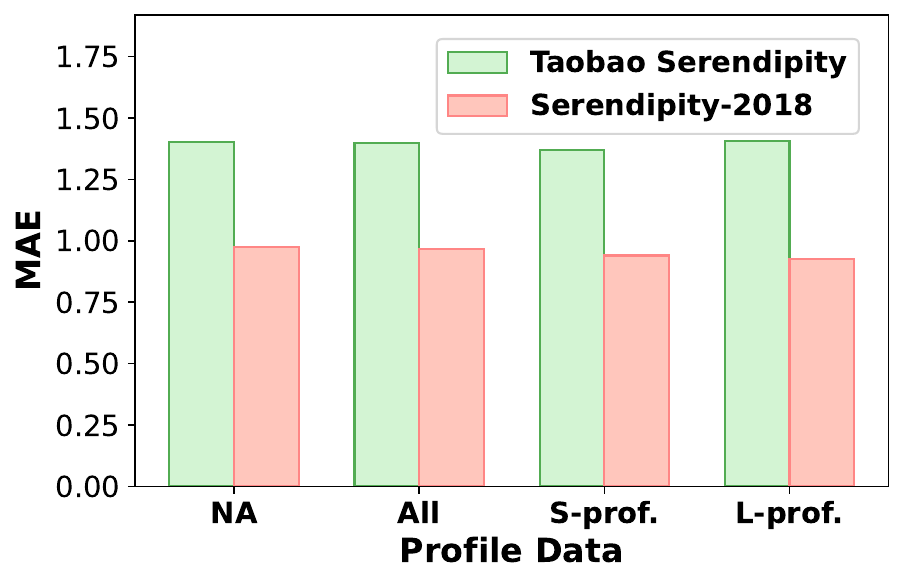}
        \caption{MAE}
        \label{fig:mae_comparison}
    \end{subfigure}
    \hfill 
    \begin{subfigure}[b]{0.32\textwidth}
        \includegraphics[width=\textwidth]{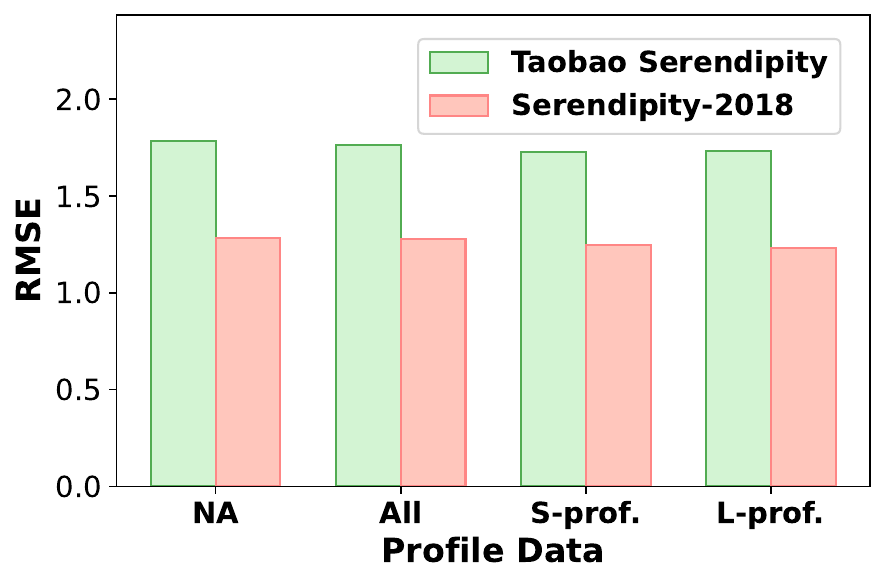}
        \caption{RMSE}
        \label{fig:rmse_comparison}
    \end{subfigure}
    \caption{Comparison w.r.t. Pearson correlation coefficient, MAE, and RMSE for different profile types (``S-prof.'' stands for short-term user profile, and ``L-prof.'' stands for long-term profile). Here, "ALL" denotes the condition with both long-term and short-term profiles.}
    \label{fig:comparison_profile}
\end{figure*} 

\begin{figure*} 
    \centering

    \begin{subfigure}[b]{0.32\textwidth}
        \includegraphics[width=\textwidth]{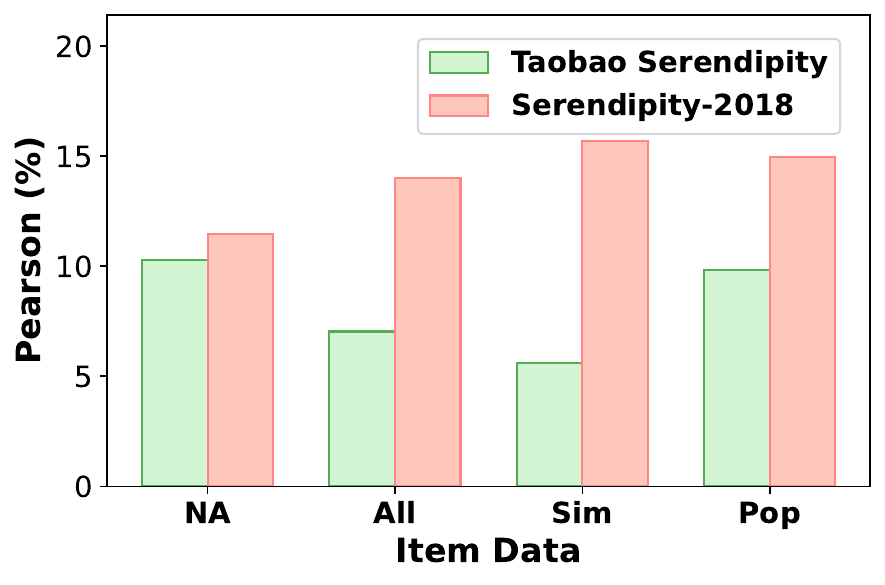}
        \caption{Pearson correlation coefficient}
        \label{fig:pearson_comparison}
    \end{subfigure}
    \hfill 
    \begin{subfigure}[b]{0.32\textwidth}
        \includegraphics[width=\textwidth]{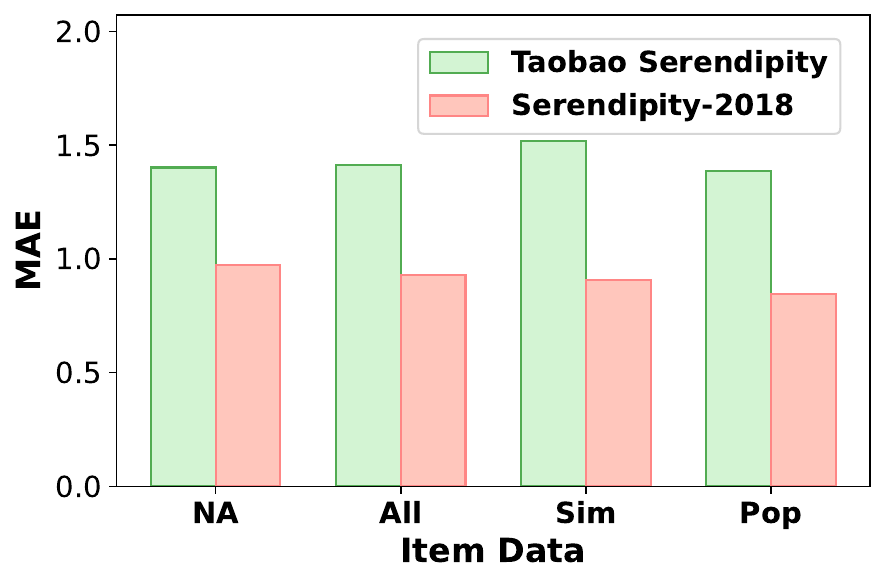}
        \caption{MAE}
        \label{fig:mae_comparison}
    \end{subfigure}
    \hfill
    \begin{subfigure}[b]{0.32\textwidth}
        \includegraphics[width=\textwidth]{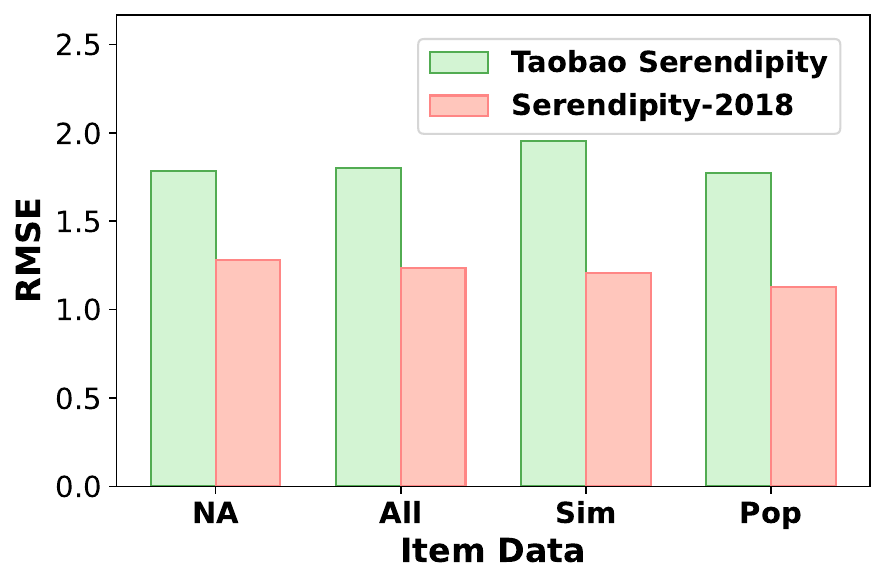}
        \caption{RMSE}
        \label{fig:rmse_comparison}
    \end{subfigure}
    \caption{Comparison w.r.t. Pearson correlation coefficient, MAE, and RMSE for different item data types (``Sim'' stands for item similarity and ``Pop'' for popularity). Here, "ALL" denotes the condition with both popularity and similarity data.}
    \label{fig:comparison_item}
\end{figure*}

\subsection{User Data}
\label{sec:user}

Since the Serendipity-2018 dataset lacks demographic and psychological user data, we mainly investigated their effects using the Taobao dataset. The results are presented in Figure~\ref{fig:comparison_demo_psy}. 
Notably, incorporating curiosity significantly improved performance, achieving a Pearson correlation coefficient of 17.83\%. This is intuitive, as curiosity influences user perceptions of serendipity: more curious users are more inclined to explore items with lower relevance but higher unexpectedness~\cite{taobao, zhao2016much, fu2023modeling}.

Conversely, attributes like the Big-Five personality traits, age, and gender did not help enhance performance and even reduced it, which contradicts previous studies linking certain Big-Five traits (e.g., extraversion and neuroticism) to serendipity~\cite{wang2023item, wang2020impacts}. We propose two possible reasons for this finding: (1) These attributes represent complex user characteristics, which current LLMs may not fully understand and simulate. (2) The relationship between these attributes and serendipity is less apparent than that of curiosity. This subtle relationship might be challenging for LLMs to capture without additional data support and instructions.

In addition, regarding user behavior history that contains the user's previous interaction data, though it can be useful for capturing user preferences, excessively long history can impair the reasoning capability of LLMs~\cite{liu2024longgenbench, li2023loogle}. 
To address this, following ~\cite{xi2025bursting}, we employ an LLM to summarize long-term interactions into a \textit{long-term user profile}, while interactions from the past 2--4 weeks are aggregated to construct the \textit{short-term profile}. 
Injecting the two types of user profiles into the LLM-based evaluator, respectively, reveals the following observation (see Figure~\ref{fig:comparison_profile}): For the Taobao dataset, incorporating the short-term profile improves the Pearson correlation coefficient. In contrast, for the Serendipity-2018 dataset, the long-term profile yields a higher Pearson correlation coefficient. The results align with the discussion from~\cite{wang2023item} that in shopping domains like Taobao, short-term relevance might play a more important role in driving serendipity, while in movie domains, long-term unexpectedness tends to have a greater influence.

\begin{figure*} 
    \centering
    \begin{subfigure}[b]{0.32\textwidth}
        \includegraphics[width=\textwidth]{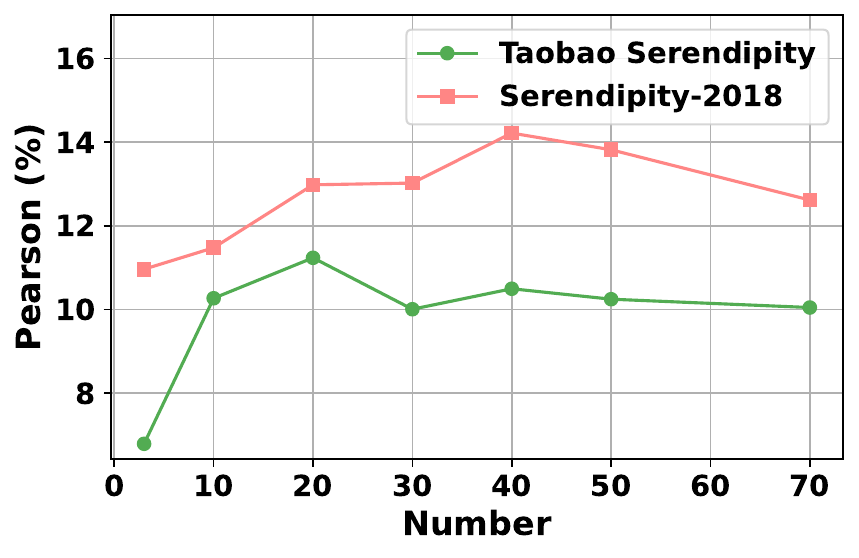}
        \caption{Pearson correlation coefficient}
        \label{fig:pearson_comparison}
    \end{subfigure}
    \hfill 
    \begin{subfigure}[b]{0.32\textwidth}
        \includegraphics[width=\textwidth]{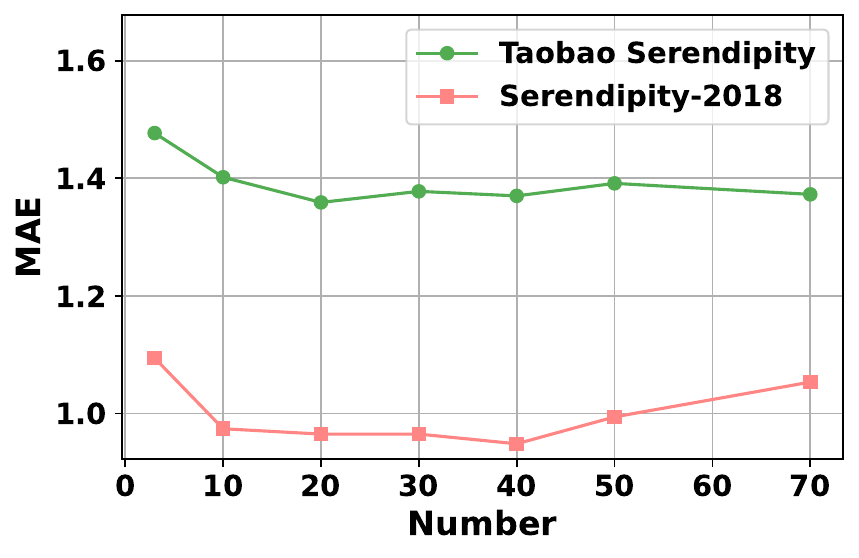}
        \caption{MAE}
        \label{fig:mae_comparison}
    \end{subfigure}
    \hfill 
    \begin{subfigure}[b]{0.32\textwidth}
        \includegraphics[width=\textwidth]{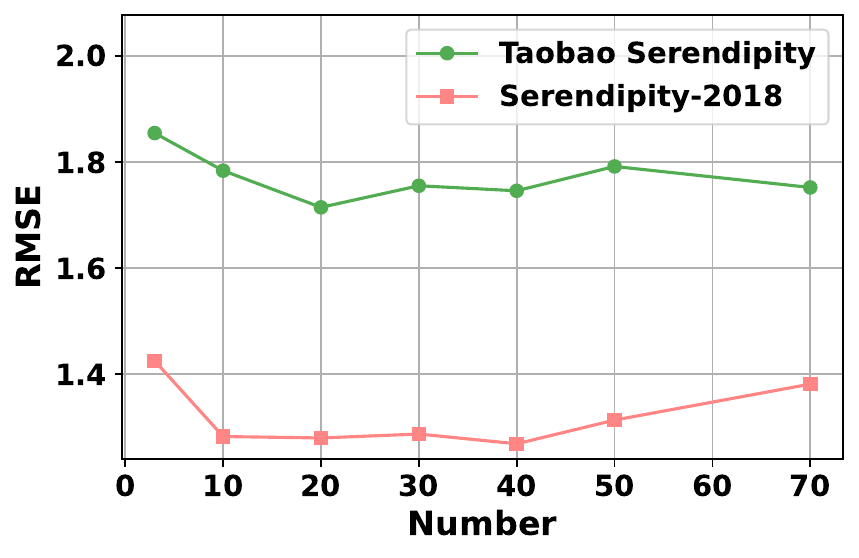}
        \caption{RMSE}
        \label{fig:rmse_comparison}
    \end{subfigure}
    \caption{Comparison w.r.t. Pearson correlation coefficient, MAE, and RMSE with varying interaction history length across the two datasets.}
    \label{fig:comparison_num}
\end{figure*}

\subsection{Item Data}
For item data, we primarily consider item popularity and similarity. Popularity is measured as the percentage of users who rated the movie in Serendipity-2018, while in Taobao, it is binary-coded (1 for items in the platform's HOT function, 0 otherwise). Similarity is represented by the minimum collaborative-based Jaccard distance between the target item and the user's historical interactions. Figure~\ref{fig:comparison_item} shows the different effects of item attributes on LLM evaluation performance across the two datasets. 

\begin{itemize}
    \item For \textit{popularity}, its incorporation yields minimal performance improvements on the Taobao dataset, but significantly enhances performance on the Serendipity-2018 dataset. This disparity stems from inherent domain differences: the movie domain naturally encourages exploration due to minimal risk and cost, making users more open to unpopular items, while e-commerce scenarios involve monetary investment that may limit such exploratory behavior.
    \item For \textit{similarity}, a similar trend is observed. On e-commerce platforms, an item that a user interacted with several months ago, even if it is very similar to the target item, might not be relevant for the current purchase decision. In contrast, for movies, user preferences tend to be more stable, and a movie with high similarity that was watched a few months ago might still have reference value. 
\end{itemize}

\begin{table}
\centering
\caption{The performance comparison regarding different interaction types. The best results are highlighted in \textbf{bold}.}
\label{tab:Interaction Type}
\small
\begin{tabular}{l l r r r}
\toprule
\textbf{Interaction Type} & \textbf{Pearson (\%)} & \textbf{MAE} & \textbf{RMSE} \\
\midrule
\multicolumn{4}{c}{\textit{Taobao Serendipity}} \\ 
\midrule
\textbf{NA} & \textbf{11.2366} & \textbf{1.3590} & \textbf{1.7145} \\
\textbf{Click} & 10.9889 & 1.3883 & 1.7831 \\ 
\textbf{Purchase} & 5.0663 & 1.3591 & 1.7375 \\
\textbf{Click \& Purchase} & 9.5309 & 1.3966 & 1.7814 \\ 
\midrule
\multicolumn{4}{c}{\textit{Serendipity-2018}} \\ 
\midrule
\textbf{NA} & 11.4769 & 0.9742 & 1.2825 \\ 
\textbf{Rating} & \textbf{12.2663} & \textbf{0.9353} & \textbf{1.2591} \\ 
\bottomrule
\end{tabular}
\end{table}

\subsection{Interaction Data}
In this section, we emphasize understanding how user-item interaction data can impact the performance of evaluation.  
\begin{itemize}
\item Regarding \textit{interaction history length}, we examined the impact of varying \textit{k} when selecting the users' top-\textit{k} most recent interactions. 
Figure~\ref{fig:comparison_num} shows that LLMs generally achieve superior performance with shorter interaction sequences. 
This phenomenon can be attributed to two possible reasons: (1) existing LLMs' limited capability in processing and reasoning over lengthy interaction sequences, and (2) more recent interactions having stronger correlation with the user's current preference and intention.
We also experimented with time-window-based filtering (e.g., interactions in recent days), but the experiments did not identify clear performance improvements. 
This might be due to the substantial variance in user interaction frequencies. Some users only recorded a few interactions within a one-day window, while others accumulated thousands over two weeks. This heterogeneity poses a significant challenge for LLM inference.


\item For \textit{interaction type}, user interaction with items can occur in different forms. For example, the Taobao dataset includes interactions like clicks and purchases, while the Serendipity-2018 dataset contains user ratings on a 1--5 scale. Due to the sparse nature of purchase behaviors, we increase the interaction history length from the default 10 to 20 for the Taobao dataset.
The results are shown in Table \ref{tab:Interaction Type}. 
It is apparent that in the Taobao dataset, excluding interaction type information produces the best results. This is plausible, as serendipity may have complex relationships with interaction types that vary among users.
Without specific fine-tuning, LLMs may struggle to capture these relationships, thereby likely leading to unsatisfactory outcomes. 
Conversely, in Serendipity-2018, incorporating rating information enhances evaluation performance. 
This may be because ratings, unlike clicks, offer more precise indicators of user preferences.
\end{itemize}
\subsection{Summary about Auxiliary Data}

Our comprehensive analysis yields two interesting findings: 
(1) Auxiliary data can improve the evaluation of recommendation serendipity by LLMs, but the effectiveness of the data type considered is dependent on the domain.
For instance, user data such as curiosity might be more beneficial in the Taobao dataset, and item data such as popularity is crucial for Serendipity-2018. 
(2) Given LLMs' current limitations in simulating human behavior~\cite{wang2025user, zhu2024reliable}, indiscriminately incorporating all available auxiliary data may not align with user study assessments~\cite{wang2023item}. This highlights the necessity and importance of our work in thoroughly investigating the roles of different types of auxiliary data for LLM-based evaluators.

\begin{figure*} 
    \centering
    \begin{subfigure}[b]{0.32\textwidth}
        \includegraphics[width=\textwidth]{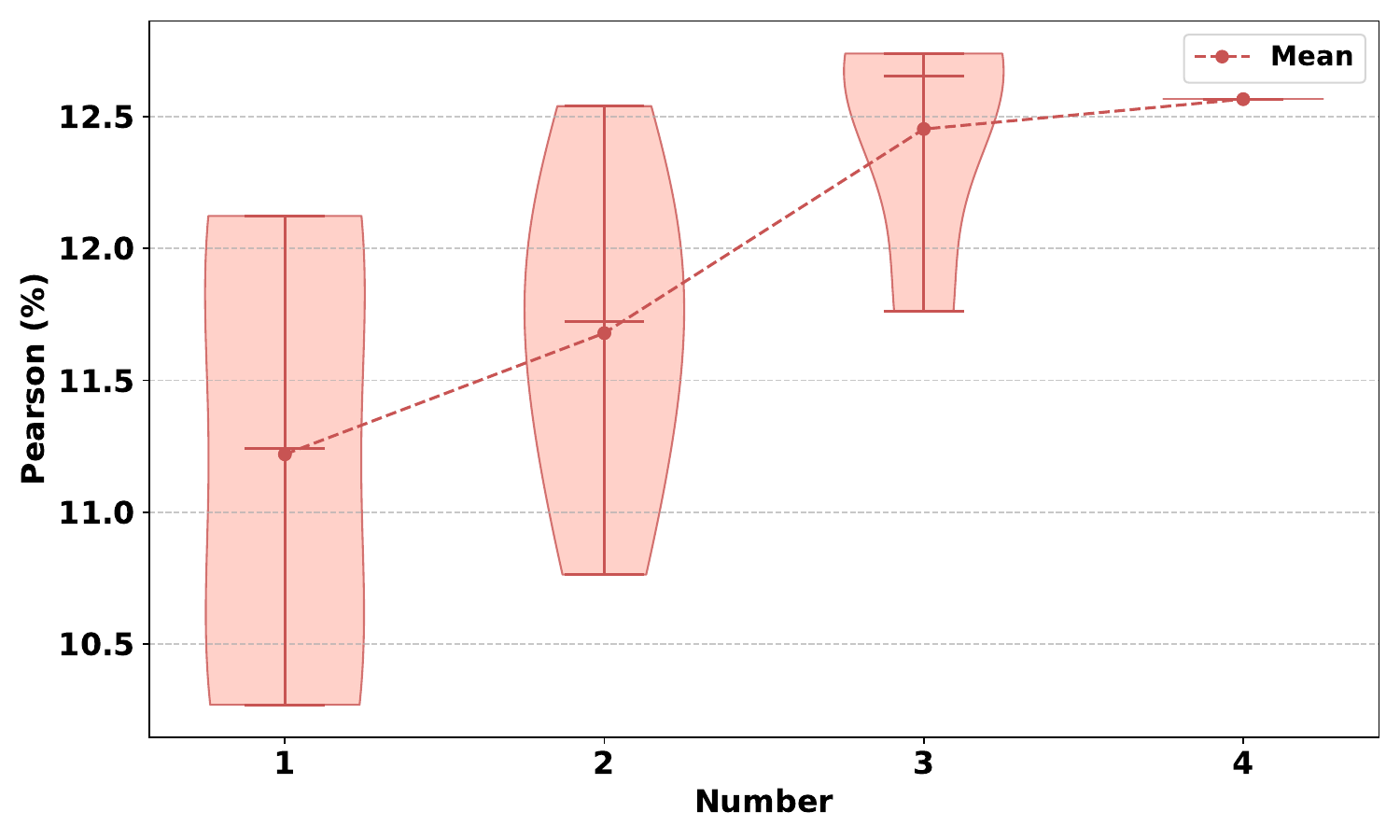}
        \caption{Pearson correlation coefficient}
        \label{fig:pearson_comparison}
    \end{subfigure}
    \hfill 
    \begin{subfigure}[b]{0.32\textwidth}
        \includegraphics[width=\textwidth]{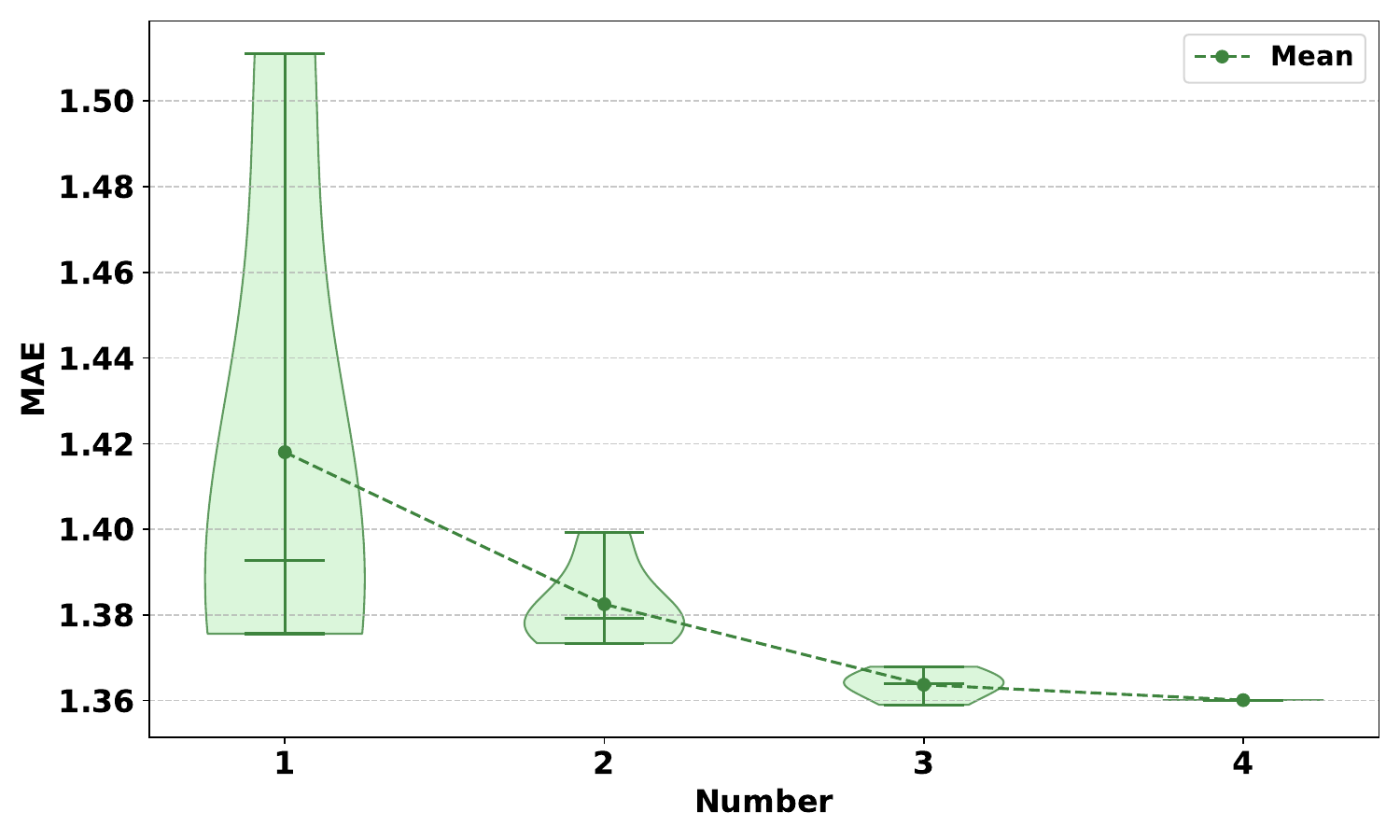}
        \caption{MAE}
        \label{fig:mae_comparison}
    \end{subfigure}
    \hfill 
    \begin{subfigure}[b]{0.32\textwidth}
        \includegraphics[width=\textwidth]{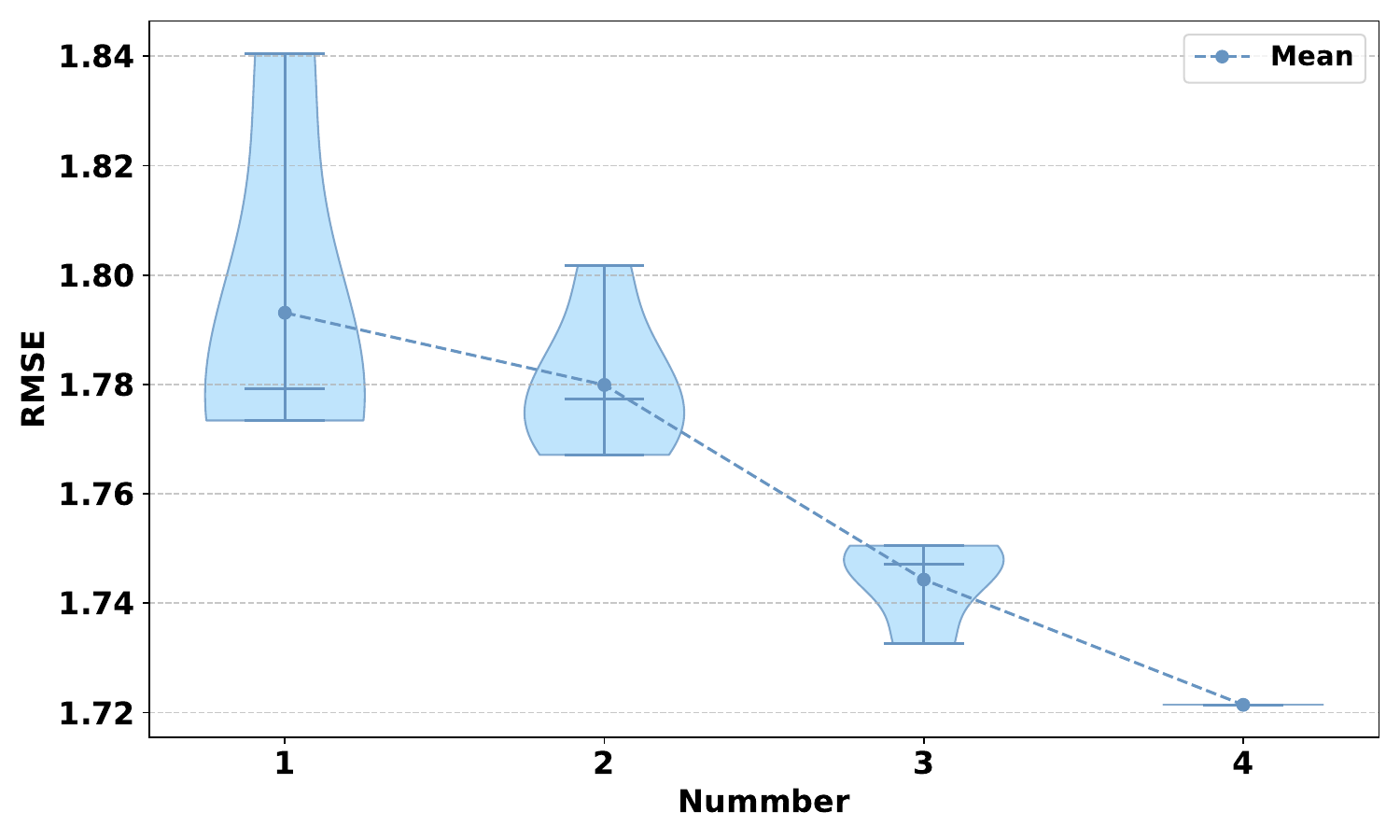}
        \caption{RMSE}
        \label{fig:rmse_comparison}
    \end{subfigure}
    \caption{Comparison w.r.t. Pearson correlation coefficient, MAE, and RMSE regarding different ensemble sizes on the Taobao.}
    \label{fig:comparison_ensem}
\end{figure*}

\begin{figure*} 
    \centering
    \begin{subfigure}[b]{0.32\textwidth}
        \includegraphics[width=\textwidth]{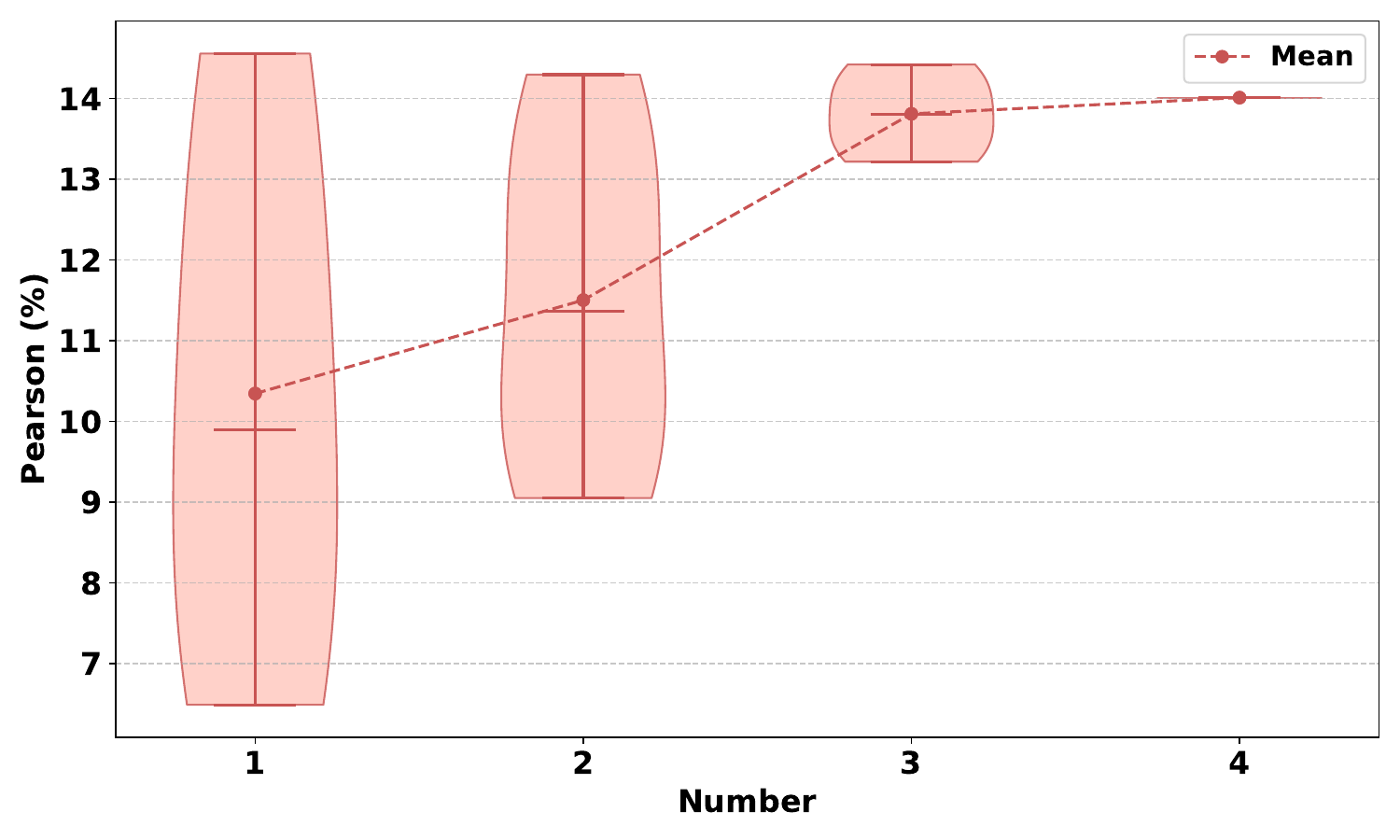}
        \caption{Pearson correlation coefficient}
        \label{fig:pearson_comparison}
    \end{subfigure}
    \hfill 
    \begin{subfigure}[b]{0.32\textwidth}
        \includegraphics[width=\textwidth]{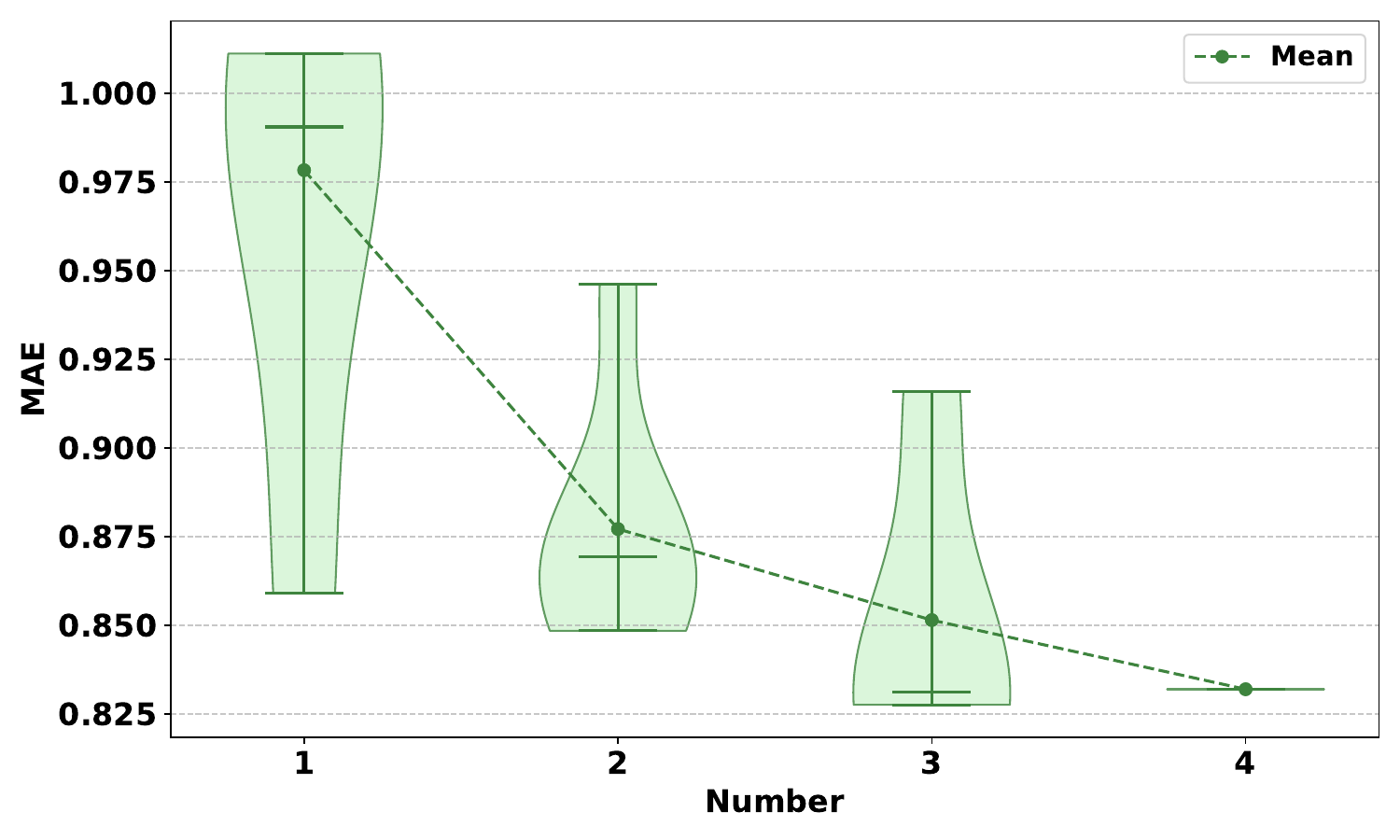}
        \caption{MAE}
        \label{fig:mae_comparison}
    \end{subfigure}
    \hfill 
    \begin{subfigure}[b]{0.32\textwidth}
        \includegraphics[width=\textwidth]{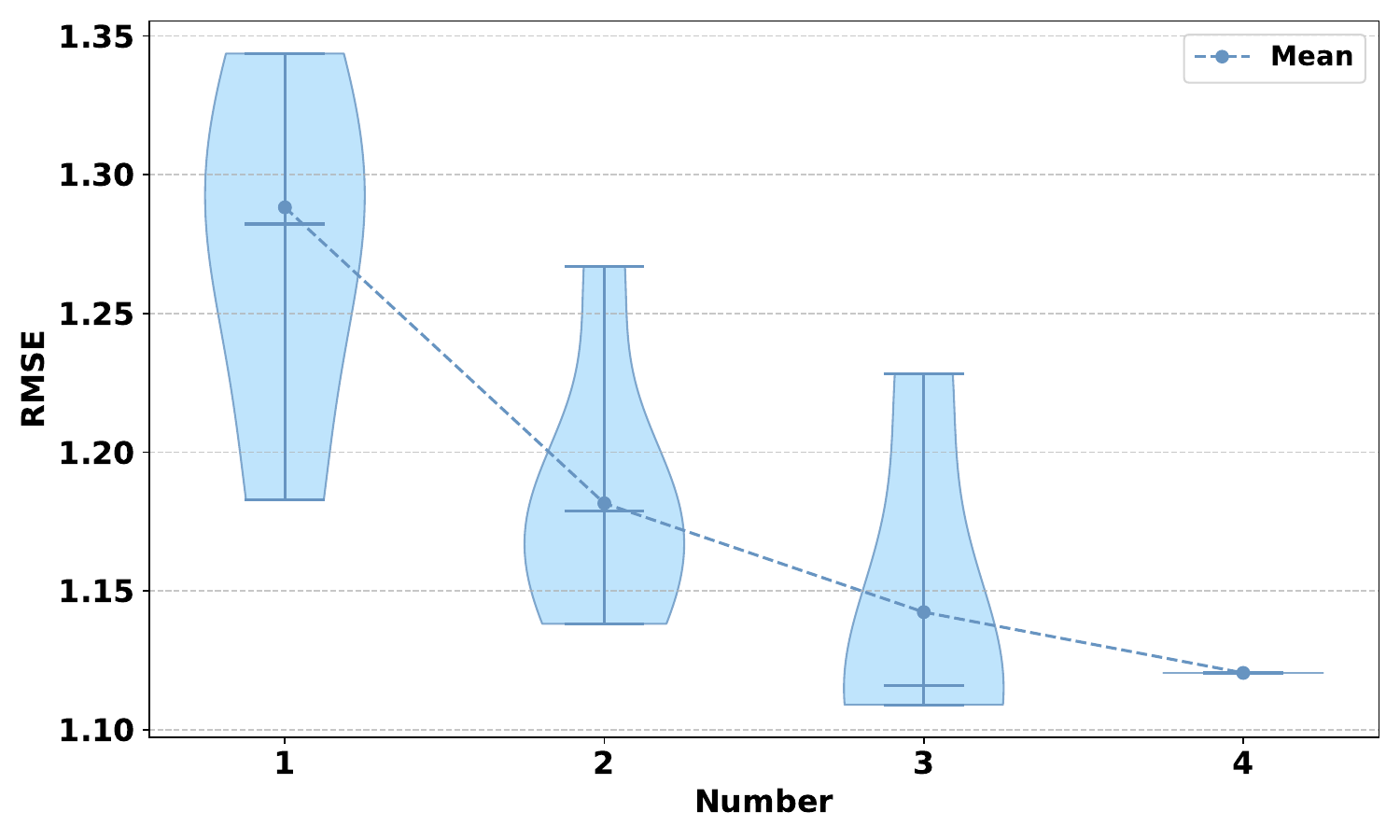}
        \caption{RMSE}
        \label{fig:rmse_comparison}
    \end{subfigure}
    \caption{Comparison w.r.t. Pearson correlation coefficient, MAE, and RMSE regarding different ensemble sizes on the Serendipity-2018 dataset.}
    \label{fig:comparison_ensem_2018}
\end{figure*}

\begin{table*} 
\centering
\caption{The performance of LLM-based serendipity evaluation by integrating multi-LLM ensemble and auxiliary data. The best results are highlighted in \textbf{bold}, and the second-best results are \underline{underlined}.}
\label{tab:summary}
{
\begin{tabular}{l|ccc|ccc}
\toprule
\multirow{2}{*}{Method} & \multicolumn{3}{c|}{Taobao Serendipity} & \multicolumn{3}{c}{Serendipity-2018} \\
\cmidrule(lr){2-4} \cmidrule(lr){5-7}
                         & Pearson(\%) & MAE & RMSE & Pearson(\%) & MAE & RMSE \\
\midrule
NA & 10.2701 & 1.4018 & 1.7836& 11.4769 & 0.9742 & 1.2825  \\
Auxiliary Data & \underline{18.2535} & \underline{1.3871} & \underline{1.7505} & \underline{20.2453} & \underline{0.8256} & \underline{1.1164}\\
Multi-LLM \& Auxiliary Data & \textbf{20.2293} & \textbf{1.2960} & \textbf{1.6238} & \textbf{21.5122} & \textbf{0.7486} & \textbf{1.0563}  \\

\bottomrule
\end{tabular}}
\end{table*}

\section{RQ3: The Power of Multi-LLM}

Multi-LLM aims to combine multiple LLMs through specific aggregation rules to determine the final output, which has demonstrated promising results in various evaluation tasks (e.g., explainability evaluation)~\cite{zhang2024large,patel2024aime}.
Therefore, in our work, we further explore the potential of multi-LLM ensembles to improve serendipity evaluation. Based on our findings from Section~\ref{sec:rq1}, we exclude models from the LLaMA family due to their previously demonstrated limitations. Specifically, we employ score averaging as the ensemble strategy, by which the final serendipity score is computed as the arithmetic mean of predictions from multiple LLMs. The results, as depicted in Figure~\ref{fig:comparison_ensem} and Figure~\ref{fig:comparison_ensem_2018}, reveal two major insights:
\begin{enumerate}
    \item An ensemble of multiple LLMs has the potential to further enhance serendipity evaluation accuracy. As the number of LLMs used increases, multi-LLM techniques show performance improvements. This suggests that a multi-LLM approach can effectively compensate for the limitations in knowledge and reasoning abilities inherent to a single LLM.
    \item The degree of improvement can be influenced by the specific combination of LLMs used in the ensemble. This is partly because some LLMs, such as those in the Qwen family, possess similar knowledge and reasoning capabilities; thus, integrating them may not yield substantial enhancements. This highlights the critical importance of strategic model selection in constructing effective ensembles.
\end{enumerate}

Furthermore, we conducted a grid search across all possible combinations by integrating Multi-LLM and auxiliary data. 
Due to space limitations, we only present the best results, as shown in Table~\ref{tab:summary}, where 
``NA'' represents the baseline condition of only using Qwen2.5-14B with the 10 most recent user interactions prior to the target item. Remarkably, we achieved impressive performance in both datasets. Specifically, the Pearson correlation coefficients reach 20.23\% and 21.51\% on the Taobao and Serendipity-2018 datasets, respectively. These optimal results obtained from the auxiliary data for the two datasets align with our previous analysis (see Section~\ref{sec:data}), and the multi-LLMs are Qwen2.5-14B, Qwen2.5-72B, and GPT-4. This impressive performance suggests that LLMs may serve as potential evaluators in the future, offering a promising balance between efficiency and effectiveness.


\section{Related Work}
Serendipity has increasingly captured the interest of the recommender systems community, due to its potential to address the information cocoon and filter bubble issues and enhance user satisfaction. 
Research has predominantly focused on developing algorithms to strengthen recommendation serendipity. 
For example, 
SerRec~\cite{senrec} introduces a novel transfer learning approach, initially using a large dataset to transfer relevance scores, which are then refined for serendipity scores using a smaller dataset. SerenCDR~\cite{SerenCDR} is pioneering in enhancing cross-domain serendipity by utilizing a deep learning model and auxiliary loss to strengthen serendipity learning within each domain. 
SerenPrompt~\cite{SerenPrompt} is the first to explore the use of prompting LLMs for serendipitous recommendations.

Despite these advancements, evaluating serendipity remains challenging due to its inherently subjective nature. 
Chen~\cite{taobao} et al. conduct an extensive user study on the Taobao platform, capturing user perceptions of serendipity directly. 
Binst~\cite{Binst} reviews current evaluation methodologies, identifying that existing metrics lack standardized and validated methods. Tokutake et al.~\cite{Japan} explore the application of LLMs in assessing serendipity. However, their study primarily focuses on validating the evaluation capabilities of LLMs and the effectiveness of the \textbf{SOG} metric. While preliminary explorations have been conducted, to our knowledge, no comprehensive and systematic study has addressed the reproducibility of various LLM techniques across diverse product domains and data types for serendipity evaluation.

\section{Conclusion}


This study investigates the potential of large language models (LLMs) for serendipity evaluation in recommender systems, a task complicated by its subjective and ambiguous nature. Through systematic meta-evaluation on two user-study-validated datasets, we show that even basic zero-shot and few-shot LLMs can match or surpass conventional proxy metrics. Incorporating auxiliary data and multi-LLM strategies further improves alignment with human judgments, with optimal configurations achieving Pearson correlation coefficients above 20\%. These findings suggest that LLM-based evaluation has the potential to combine the accuracy of user studies with the efficiency of proxy metrics, offering a promising direction for serendipity evaluation in recommender systems. In the future, we aim to explore more advanced LLM techniques and investigate the explainability of serendipity.
\begin{acks}
This work is supported by Hong Kong Baptist University IG-FNRA Project (RC-FNRA-IG/21-22/SCI/01), Key Research Partnership Scheme (KRPS/23-24/02), and NSFC/RGC Joint Research Scheme (N\_HKBU214/24).
\end{acks}

\clearpage
\balance
\bibliographystyle{ACM-Reference-Format}
\bibliography{paper}


\begin{thebibliography}{43}


\ifx \showCODEN    \undefined \def \showCODEN     #1{\unskip}     \fi
\ifx \showISBNx    \undefined \def \showISBNx     #1{\unskip}     \fi
\ifx \showISBNxiii \undefined \def \showISBNxiii  #1{\unskip}     \fi
\ifx \showISSN     \undefined \def \showISSN      #1{\unskip}     \fi
\ifx \showLCCN     \undefined \def \showLCCN      #1{\unskip}     \fi
\ifx \shownote     \undefined \def \shownote      #1{#1}          \fi
\ifx \showarticletitle \undefined \def \showarticletitle #1{#1}   \fi
\ifx \showURL      \undefined \def \showURL       {\relax}        \fi
\providecommand\bibfield[2]{#2}
\providecommand\bibinfo[2]{#2}
\providecommand\natexlab[1]{#1}
\providecommand\showeprint[2][]{arXiv:#2}

\bibitem[Achiam et~al\mbox{.}(2023)]%
        {achiam2023gpt}
\bibfield{author}{\bibinfo{person}{Josh Achiam}, \bibinfo{person}{Steven Adler}, {et~al\mbox{.}}} \bibinfo{year}{2023}\natexlab{}.
\newblock \showarticletitle{GPT-4 Technical Report}.
\newblock \bibinfo{journal}{\emph{arXiv}} (\bibinfo{year}{2023}).
\newblock


\bibitem[Ashktorab et~al\mbox{.}(2024)]%
        {ashktorab2024aligning}
\bibfield{author}{\bibinfo{person}{Zahra Ashktorab}, \bibinfo{person}{Michael Desmond}, \bibinfo{person}{Qian Pan}, \bibinfo{person}{James~M Johnson}, \bibinfo{person}{Martin~Santillan Cooper}, \bibinfo{person}{Elizabeth~M Daly}, \bibinfo{person}{Rahul Nair}, \bibinfo{person}{Tejaswini Pedapati}, \bibinfo{person}{Swapnaja Achintalwar}, {and} \bibinfo{person}{Werner Geyer}.} \bibinfo{year}{2024}\natexlab{}.
\newblock \showarticletitle{Aligning Human and LLM Judgments: Insights from EvalAssist on Task-Specific Evaluations and AI-assisted Assessment Strategy Preferences}.
\newblock \bibinfo{journal}{\emph{arXiv preprint arXiv:2410.00873}} (\bibinfo{year}{2024}).
\newblock


\bibitem[Bavaresco et~al\mbox{.}(2024)]%
        {bavaresco2024llms}
\bibfield{author}{\bibinfo{person}{Anna Bavaresco}, \bibinfo{person}{Raffaella Bernardi}, \bibinfo{person}{Leonardo Bertolazzi}, \bibinfo{person}{Desmond Elliott}, \bibinfo{person}{Raquel Fern{\'a}ndez}, \bibinfo{person}{Albert Gatt}, \bibinfo{person}{Esam Ghaleb}, \bibinfo{person}{Mario Giulianelli}, \bibinfo{person}{Michael Hanna}, \bibinfo{person}{Alexander Koller}, {et~al\mbox{.}}} \bibinfo{year}{2024}\natexlab{}.
\newblock \showarticletitle{LLMs Instead of Human Judges? A Large-Scale Empirical Study Across 20 NLP Evaluation Tasks}.
\newblock \bibinfo{journal}{\emph{arXiv preprint arXiv:2406.18403}} (\bibinfo{year}{2024}).
\newblock


\bibitem[Binst(2024)]%
        {Binst}
\bibfield{author}{\bibinfo{person}{Brett Binst}.} \bibinfo{year}{2024}\natexlab{}.
\newblock \showarticletitle{How to Evaluate Serendipity in Recommender Systems: the Need for a Serendiptionnaire}. In \bibinfo{booktitle}{\emph{RecSys}}. \bibinfo{pages}{1335--1341}.
\newblock


\bibitem[Chen et~al\mbox{.}(2019)]%
        {taobao}
\bibfield{author}{\bibinfo{person}{Li Chen}, \bibinfo{person}{Yonghua Yang}, \bibinfo{person}{Ningxia Wang}, \bibinfo{person}{Keping Yang}, {and} \bibinfo{person}{Quan Yuan}.} \bibinfo{year}{2019}\natexlab{}.
\newblock \showarticletitle{How Serendipity Improves User Satisfaction with Recommendations? A Large-Scale User Evaluation}. In \bibinfo{booktitle}{\emph{WWW}}. \bibinfo{pages}{240--250}.
\newblock


\bibitem[Fu and Niu(2023)]%
        {fu2023modeling}
\bibfield{author}{\bibinfo{person}{Zhe Fu} {and} \bibinfo{person}{Xi Niu}.} \bibinfo{year}{2023}\natexlab{}.
\newblock \showarticletitle{Modeling Users’ Curiosity in Recommender Systems}.
\newblock \bibinfo{journal}{\emph{TKDD}} \bibinfo{volume}{18}, \bibinfo{number}{1} (\bibinfo{year}{2023}), \bibinfo{pages}{1--23}.
\newblock


\bibitem[Fu and Niu(2024)]%
        {SerenPrompt}
\bibfield{author}{\bibinfo{person}{Zhe Fu} {and} \bibinfo{person}{Xi Niu}.} \bibinfo{year}{2024}\natexlab{}.
\newblock \showarticletitle{The Art of Asking: Prompting Large Language Models for Serendipity Recommendations}. In \bibinfo{booktitle}{\emph{ICTIR}}. \bibinfo{pages}{157--166}.
\newblock


\bibitem[Fu et~al\mbox{.}(2023a)]%
        {fu2023deep}
\bibfield{author}{\bibinfo{person}{Zhe Fu}, \bibinfo{person}{Xi Niu}, {and} \bibinfo{person}{Mary~Lou Maher}.} \bibinfo{year}{2023}\natexlab{a}.
\newblock \showarticletitle{Deep Learning Models for Serendipity Recommendations: A Survey and New Perspectives}.
\newblock \bibinfo{journal}{\emph{Comput. Surveys}} \bibinfo{volume}{56}, \bibinfo{number}{1} (\bibinfo{year}{2023}), \bibinfo{pages}{1--26}.
\newblock


\bibitem[Fu et~al\mbox{.}(2025)]%
        {SerenCDR}
\bibfield{author}{\bibinfo{person}{Zhe Fu}, \bibinfo{person}{Xi Niu}, \bibinfo{person}{Xiangcheng Wu}, {and} \bibinfo{person}{Ruhani Rahman}.} \bibinfo{year}{2025}\natexlab{}.
\newblock \showarticletitle{A Deep Learning Model for Cross-Domain Serendipity Recommendations}.
\newblock \bibinfo{journal}{\emph{TORS}} \bibinfo{volume}{3}, \bibinfo{number}{3} (\bibinfo{year}{2025}), \bibinfo{pages}{1--21}.
\newblock


\bibitem[Fu et~al\mbox{.}(2023b)]%
        {fu2023wisdom}
\bibfield{author}{\bibinfo{person}{Zhe Fu}, \bibinfo{person}{Xi Niu}, {and} \bibinfo{person}{Li Yu}.} \bibinfo{year}{2023}\natexlab{b}.
\newblock \showarticletitle{Wisdom of Crowds and Fine-grained Learning for Serendipity Recommendations}. In \bibinfo{booktitle}{\emph{SIGIR}}. \bibinfo{pages}{739--748}.
\newblock


\bibitem[Gosling et~al\mbox{.}(2003)]%
        {gosling2003very}
\bibfield{author}{\bibinfo{person}{Samuel~D Gosling}, \bibinfo{person}{Peter~J Rentfrow}, {and} \bibinfo{person}{William~B Swann~Jr}.} \bibinfo{year}{2003}\natexlab{}.
\newblock \showarticletitle{A Very Brief Measure of the Big-Five Personality Domains}.
\newblock \bibinfo{journal}{\emph{Journal of Research in personality}} \bibinfo{volume}{37}, \bibinfo{number}{6} (\bibinfo{year}{2003}), \bibinfo{pages}{504--528}.
\newblock


\bibitem[Hasan and Bunescu(2023)]%
        {hasan2023topic}
\bibfield{author}{\bibinfo{person}{Tonmoy Hasan} {and} \bibinfo{person}{Razvan Bunescu}.} \bibinfo{year}{2023}\natexlab{}.
\newblock \showarticletitle{Topic-Level Bayesian Surprise and Serendipity for Recommender Systems}. In \bibinfo{booktitle}{\emph{RecSys}}. \bibinfo{pages}{933--939}.
\newblock


\bibitem[Kashdan et~al\mbox{.}(2009)]%
        {kashdan2009curiosity}
\bibfield{author}{\bibinfo{person}{Todd~B Kashdan}, \bibinfo{person}{Matthew~W Gallagher}, \bibinfo{person}{Paul~J Silvia}, \bibinfo{person}{Beate~P Winterstein}, \bibinfo{person}{William~E Breen}, \bibinfo{person}{Daniel Terhar}, {and} \bibinfo{person}{Michael~F Steger}.} \bibinfo{year}{2009}\natexlab{}.
\newblock \showarticletitle{The Curiosity And Exploration Inventory-II: Development, Factor Structure, And Psychometrics}.
\newblock \bibinfo{journal}{\emph{Journal of research in personality}} \bibinfo{volume}{43}, \bibinfo{number}{6} (\bibinfo{year}{2009}), \bibinfo{pages}{987--998}.
\newblock


\bibitem[Kotkov et~al\mbox{.}(2018)]%
        {seren2018}
\bibfield{author}{\bibinfo{person}{Denis Kotkov}, \bibinfo{person}{Joseph~A Konstan}, \bibinfo{person}{Qian Zhao}, {and} \bibinfo{person}{Jari Veijalainen}.} \bibinfo{year}{2018}\natexlab{}.
\newblock \showarticletitle{Investigating Serendipity in Recommender Systems Based on Real User Feedback}. In \bibinfo{booktitle}{\emph{SAC}}. \bibinfo{pages}{1341--1350}.
\newblock


\bibitem[Kotkov et~al\mbox{.}(2023)]%
        {kotkov2023rethinking}
\bibfield{author}{\bibinfo{person}{Denis Kotkov}, \bibinfo{person}{Alan Medlar}, {and} \bibinfo{person}{Dorota Glowacka}.} \bibinfo{year}{2023}\natexlab{}.
\newblock \showarticletitle{Rethinking Serendipity in Recommender Systems}. In \bibinfo{booktitle}{\emph{CHIIR}}. \bibinfo{pages}{383--387}.
\newblock


\bibitem[Kotkov et~al\mbox{.}(2024)]%
        {kotkov2024dark}
\bibfield{author}{\bibinfo{person}{Denis Kotkov}, \bibinfo{person}{Alan Medlar}, \bibinfo{person}{Triin Kask}, {and} \bibinfo{person}{Dorota Glowacka}.} \bibinfo{year}{2024}\natexlab{}.
\newblock \showarticletitle{The dark matter of serendipity in recommender systems}. In \bibinfo{booktitle}{\emph{CHIIR}}. \bibinfo{pages}{108--118}.
\newblock


\bibitem[Kotkov et~al\mbox{.}(2020)]%
        {kotkov2020does}
\bibfield{author}{\bibinfo{person}{Denis Kotkov}, \bibinfo{person}{Jari Veijalainen}, {and} \bibinfo{person}{Shuaiqiang Wang}.} \bibinfo{year}{2020}\natexlab{}.
\newblock \showarticletitle{How Does Serendipity Affect Diversity in Recommender Systems? A Serendipity-Oriented Greedy Algorithm}.
\newblock \bibinfo{journal}{\emph{Computing}}  \bibinfo{volume}{102} (\bibinfo{year}{2020}), \bibinfo{pages}{393--411}.
\newblock


\bibitem[Li et~al\mbox{.}(2024)]%
        {li2024llms}
\bibfield{author}{\bibinfo{person}{Haitao Li}, \bibinfo{person}{Qian Dong}, \bibinfo{person}{Junjie Chen}, \bibinfo{person}{Huixue Su}, \bibinfo{person}{Yujia Zhou}, \bibinfo{person}{Qingyao Ai}, \bibinfo{person}{Ziyi Ye}, {and} \bibinfo{person}{Yiqun Liu}.} \bibinfo{year}{2024}\natexlab{}.
\newblock \showarticletitle{LLMs-as-Judges: A Comprehensive Survey on LLM-Based Evaluation Methods}.
\newblock \bibinfo{journal}{\emph{arXiv preprint arXiv:2412.05579}} (\bibinfo{year}{2024}).
\newblock


\bibitem[Li et~al\mbox{.}(2023)]%
        {li2023loogle}
\bibfield{author}{\bibinfo{person}{Jiaqi Li}, \bibinfo{person}{Mengmeng Wang}, \bibinfo{person}{Zilong Zheng}, {and} \bibinfo{person}{Muhan Zhang}.} \bibinfo{year}{2023}\natexlab{}.
\newblock \showarticletitle{LooGLE: Can Long-Context Language Models Understand Long Contexts?}
\newblock \bibinfo{journal}{\emph{arXiv preprint arXiv:2311.04939}} (\bibinfo{year}{2023}).
\newblock


\bibitem[Li et~al\mbox{.}(2020b)]%
        {PURS}
\bibfield{author}{\bibinfo{person}{Pan Li}, \bibinfo{person}{Maofei Que}, \bibinfo{person}{Zhichao Jiang}, \bibinfo{person}{Yao Hu}, {and} \bibinfo{person}{Alexander Tuzhilin}.} \bibinfo{year}{2020}\natexlab{b}.
\newblock \showarticletitle{PURS: Personalized Unexpected Recommender System for Improving User Satisfaction}. In \bibinfo{booktitle}{\emph{RecSys}}. \bibinfo{pages}{279--288}.
\newblock


\bibitem[Li et~al\mbox{.}(2020a)]%
        {DESR}
\bibfield{author}{\bibinfo{person}{Xueqi Li}, \bibinfo{person}{Wenjun Jiang}, \bibinfo{person}{Weiguang Chen}, \bibinfo{person}{Jie Wu}, \bibinfo{person}{Guojun Wang}, {and} \bibinfo{person}{Kenli Li}.} \bibinfo{year}{2020}\natexlab{a}.
\newblock \showarticletitle{Directional and Explainable Serendipity Recommendation}. In \bibinfo{booktitle}{\emph{WWW}}. \bibinfo{pages}{122--132}.
\newblock


\bibitem[Liu et~al\mbox{.}(2024)]%
        {liu2024longgenbench}
\bibfield{author}{\bibinfo{person}{Xiang Liu}, \bibinfo{person}{Peijie Dong}, \bibinfo{person}{Xuming Hu}, {and} \bibinfo{person}{Xiaowen Chu}.} \bibinfo{year}{2024}\natexlab{}.
\newblock \showarticletitle{Longgenbench: Long-context Generation Benchmark}.
\newblock \bibinfo{journal}{\emph{arXiv preprint arXiv:2410.04199}} (\bibinfo{year}{2024}).
\newblock


\bibitem[Pandey et~al\mbox{.}(2018)]%
        {senrec}
\bibfield{author}{\bibinfo{person}{Gaurav Pandey}, \bibinfo{person}{Denis Kotkov}, {and} \bibinfo{person}{Alexander Semenov}.} \bibinfo{year}{2018}\natexlab{}.
\newblock \showarticletitle{Recommending Serendipitous Items Using Transfer Learning}. In \bibinfo{booktitle}{\emph{CIKM}}. \bibinfo{pages}{1771--1774}.
\newblock


\bibitem[Patel et~al\mbox{.}(2024)]%
        {patel2024aime}
\bibfield{author}{\bibinfo{person}{Bhrij Patel}, \bibinfo{person}{Souradip Chakraborty}, \bibinfo{person}{Wesley~A Suttle}, \bibinfo{person}{Mengdi Wang}, \bibinfo{person}{Amrit~Singh Bedi}, {and} \bibinfo{person}{Dinesh Manocha}.} \bibinfo{year}{2024}\natexlab{}.
\newblock \showarticletitle{AIME: AI System Optimization via Multiple LLM Evaluators}.
\newblock \bibinfo{journal}{\emph{arXiv preprint arXiv:2410.03131}} (\bibinfo{year}{2024}).
\newblock


\bibitem[Pu et~al\mbox{.}(2011)]%
        {pu2011user}
\bibfield{author}{\bibinfo{person}{Pearl Pu}, \bibinfo{person}{Li Chen}, {and} \bibinfo{person}{Rong Hu}.} \bibinfo{year}{2011}\natexlab{}.
\newblock \showarticletitle{A User-Centric Evaluation Framework for Recommender Systems}. In \bibinfo{booktitle}{\emph{Proceedings of the fifth ACM conference on Recommender systems}}. \bibinfo{pages}{157--164}.
\newblock


\bibitem[Tokutake and Okamoto(2024)]%
        {Japan}
\bibfield{author}{\bibinfo{person}{Yu Tokutake} {and} \bibinfo{person}{Kazushi Okamoto}.} \bibinfo{year}{2024}\natexlab{}.
\newblock \showarticletitle{Can Large Language Models Assess Serendipity in Recommender Systems?}
\newblock \bibinfo{journal}{\emph{JACIII}} \bibinfo{volume}{28}, \bibinfo{number}{6} (\bibinfo{year}{2024}), \bibinfo{pages}{1263--1272}.
\newblock


\bibitem[Touvron et~al\mbox{.}(2023)]%
        {llama2}
\bibfield{author}{\bibinfo{person}{Hugo Touvron}, \bibinfo{person}{Louis Martin}, \bibinfo{person}{Kevin Stone}, \bibinfo{person}{Peter Albert}, \bibinfo{person}{Amjad Almahairi}, \bibinfo{person}{Yasmine Babaei}, \bibinfo{person}{Nikolay Bashlykov}, \bibinfo{person}{Soumya Batra}, \bibinfo{person}{Prajjwal Bhargava}, \bibinfo{person}{Shruti Bhosale}, {et~al\mbox{.}}} \bibinfo{year}{2023}\natexlab{}.
\newblock \showarticletitle{Llama 2: Open Foundation and Fine-Tuned Chat Models}.
\newblock \bibinfo{journal}{\emph{arXiv preprint arXiv:2307.09288}} (\bibinfo{year}{2023}).
\newblock


\bibitem[Tseng et~al\mbox{.}(2024)]%
        {tseng2024expert}
\bibfield{author}{\bibinfo{person}{Yu-Min Tseng}, \bibinfo{person}{Wei-Lin Chen}, \bibinfo{person}{Chung-Chi Chen}, {and} \bibinfo{person}{Hsin-Hsi Chen}.} \bibinfo{year}{2024}\natexlab{}.
\newblock \showarticletitle{Are Expert-Level Language Models Expert-Level Annotators?}
\newblock \bibinfo{journal}{\emph{arXiv preprint arXiv:2410.03254}} (\bibinfo{year}{2024}).
\newblock


\bibitem[Wang et~al\mbox{.}(2024)]%
        {wang2024llms}
\bibfield{author}{\bibinfo{person}{Jianling Wang}, \bibinfo{person}{Haokai Lu}, \bibinfo{person}{Yifan Liu}, \bibinfo{person}{He Ma}, \bibinfo{person}{Yueqi Wang}, \bibinfo{person}{Yang Gu}, \bibinfo{person}{Shuzhou Zhang}, \bibinfo{person}{Ningren Han}, \bibinfo{person}{Shuchao Bi}, \bibinfo{person}{Lexi Baugher}, {et~al\mbox{.}}} \bibinfo{year}{2024}\natexlab{}.
\newblock \showarticletitle{LLMs for User Interest Exploration in Large-Scale Recommendation Systems}. In \bibinfo{booktitle}{\emph{RecSys}}. \bibinfo{pages}{872--877}.
\newblock


\bibitem[Wang et~al\mbox{.}(2025)]%
        {wang2025user}
\bibfield{author}{\bibinfo{person}{Lei Wang}, \bibinfo{person}{Jingsen Zhang}, \bibinfo{person}{Hao Yang}, \bibinfo{person}{Zhi-Yuan Chen}, \bibinfo{person}{Jiakai Tang}, \bibinfo{person}{Zeyu Zhang}, \bibinfo{person}{Xu Chen}, \bibinfo{person}{Yankai Lin}, \bibinfo{person}{Hao Sun}, \bibinfo{person}{Ruihua Song}, {et~al\mbox{.}}} \bibinfo{year}{2025}\natexlab{}.
\newblock \showarticletitle{User Behavior Simulation with Large Language Model-Based Agents}.
\newblock \bibinfo{journal}{\emph{TOIS}} \bibinfo{volume}{43}, \bibinfo{number}{2} (\bibinfo{year}{2025}), \bibinfo{pages}{1--37}.
\newblock


\bibitem[Wang and Chen(2023)]%
        {wang2023item}
\bibfield{author}{\bibinfo{person}{Ningxia Wang} {and} \bibinfo{person}{Li Chen}.} \bibinfo{year}{2023}\natexlab{}.
\newblock \showarticletitle{How Do Item Features and User Characteristics Affect Users’ Perceptions of Recommendation Serendipity? A Cross-Domain Analysis}.
\newblock \bibinfo{journal}{\emph{UMUAI}} \bibinfo{volume}{33}, \bibinfo{number}{3} (\bibinfo{year}{2023}), \bibinfo{pages}{727--765}.
\newblock


\bibitem[Wang et~al\mbox{.}(2020)]%
        {wang2020impacts}
\bibfield{author}{\bibinfo{person}{Ningxia Wang}, \bibinfo{person}{Li Chen}, {et~al\mbox{.}}} \bibinfo{year}{2020}\natexlab{}.
\newblock \showarticletitle{The Impacts of Item Features and User Characteristics on Users' Perceived Serendipity of Recommendations}. In \bibinfo{booktitle}{\emph{UMAP}}.
\newblock


\bibitem[Wendler et~al\mbox{.}(2024)]%
        {wendler2024llamas}
\bibfield{author}{\bibinfo{person}{Chris Wendler}, \bibinfo{person}{Veniamin Veselovsky}, \bibinfo{person}{Giovanni Monea}, {and} \bibinfo{person}{Robert West}.} \bibinfo{year}{2024}\natexlab{}.
\newblock \showarticletitle{Do Llamas Work in English? On the Latent Language of Multilingual Transformers}. In \bibinfo{booktitle}{\emph{ACL}}. \bibinfo{pages}{15366--15394}.
\newblock


\bibitem[Xi et~al\mbox{.}(2025)]%
        {xi2025bursting}
\bibfield{author}{\bibinfo{person}{Yunjia Xi}, \bibinfo{person}{Muyan Weng}, \bibinfo{person}{Wen Chen}, \bibinfo{person}{Chao Yi}, \bibinfo{person}{Dian Chen}, \bibinfo{person}{Gaoyang Guo}, \bibinfo{person}{Mao Zhang}, \bibinfo{person}{Jian Wu}, \bibinfo{person}{Yuning Jiang}, \bibinfo{person}{Qingwen Liu}, {et~al\mbox{.}}} \bibinfo{year}{2025}\natexlab{}.
\newblock \showarticletitle{Bursting Filter Bubble: Enhancing Serendipity Recommendations with Aligned Large Language Models}.
\newblock \bibinfo{journal}{\emph{arXiv preprint arXiv:2502.13539}} (\bibinfo{year}{2025}).
\newblock


\bibitem[Yang et~al\mbox{.}(2024)]%
        {yang2024qwen2}
\bibfield{author}{\bibinfo{person}{An Yang}, \bibinfo{person}{Baosong Yang}, \bibinfo{person}{Beichen Zhang}, \bibinfo{person}{Binyuan Hui}, \bibinfo{person}{Bo Zheng}, \bibinfo{person}{Bowen Yu}, \bibinfo{person}{Chengyuan Li}, \bibinfo{person}{Dayiheng Liu}, \bibinfo{person}{Fei Huang}, \bibinfo{person}{Haoran Wei}, {et~al\mbox{.}}} \bibinfo{year}{2024}\natexlab{}.
\newblock \showarticletitle{Qwen2.5 Technical Report}.
\newblock \bibinfo{journal}{\emph{arXiv preprint arXiv:2412.15115}} (\bibinfo{year}{2024}).
\newblock


\bibitem[Zhang et~al\mbox{.}(2021)]%
        {SNPR}
\bibfield{author}{\bibinfo{person}{Mingwei Zhang}, \bibinfo{person}{Yang Yang}, \bibinfo{person}{Rizwan Abbas}, \bibinfo{person}{Ke Deng}, \bibinfo{person}{Jianxin Li}, {and} \bibinfo{person}{Bin Zhang}.} \bibinfo{year}{2021}\natexlab{}.
\newblock \showarticletitle{SNPR: A Serendipity-Oriented Next POI Recommendation Model}. In \bibinfo{booktitle}{\emph{CIKM}}. \bibinfo{pages}{2568--2577}.
\newblock


\bibitem[Zhang et~al\mbox{.}(2024)]%
        {zhang2024large}
\bibfield{author}{\bibinfo{person}{Xiaoyu Zhang}, \bibinfo{person}{Yishan Li}, \bibinfo{person}{Jiayin Wang}, \bibinfo{person}{Bowen Sun}, \bibinfo{person}{Weizhi Ma}, \bibinfo{person}{Peijie Sun}, {and} \bibinfo{person}{Min Zhang}.} \bibinfo{year}{2024}\natexlab{}.
\newblock \showarticletitle{Large Language Models as Evaluators for Recommendation Explanations}. In \bibinfo{booktitle}{\emph{RecSys}}. \bibinfo{pages}{33--42}.
\newblock


\bibitem[Zhao et~al\mbox{.}(2024b)]%
        {zhao2024llama}
\bibfield{author}{\bibinfo{person}{Jun Zhao}, \bibinfo{person}{Zhihao Zhang}, \bibinfo{person}{Luhui Gao}, \bibinfo{person}{Qi Zhang}, \bibinfo{person}{Tao Gui}, {and} \bibinfo{person}{Xuanjing Huang}.} \bibinfo{year}{2024}\natexlab{b}.
\newblock \showarticletitle{Llama Beyond English: An Empirical Study on Language Capability Transfer}.
\newblock \bibinfo{journal}{\emph{arXiv preprint arXiv:2401.01055}} (\bibinfo{year}{2024}).
\newblock


\bibitem[Zhao and Lee(2016)]%
        {zhao2016much}
\bibfield{author}{\bibinfo{person}{Pengfei Zhao} {and} \bibinfo{person}{Dik~Lun Lee}.} \bibinfo{year}{2016}\natexlab{}.
\newblock \showarticletitle{How Much Novelty Is Relevant? It Depends on Your Curiosity}. In \bibinfo{booktitle}{\emph{SIGIR}}. \bibinfo{pages}{315--324}.
\newblock


\bibitem[Zhao et~al\mbox{.}(2025)]%
        {ZCC25}
\bibfield{author}{\bibinfo{person}{Yuhan Zhao}, \bibinfo{person}{Rui Chen}, \bibinfo{person}{Li Chen}, \bibinfo{person}{Shuang Zhang}, \bibinfo{person}{Qilong Han}, {and} \bibinfo{person}{Hongtao Song}.} \bibinfo{year}{2025}\natexlab{}.
\newblock \showarticletitle{From Pairwise to Ranking: Climbing the Ladder to Ideal Collaborative Filtering with Pseudo-Ranking}. In \bibinfo{booktitle}{\emph{AAAI}}, Vol.~\bibinfo{volume}{39}. \bibinfo{pages}{13392--13400}.
\newblock


\bibitem[Zhao et~al\mbox{.}(2024a)]%
        {ZCH24}
\bibfield{author}{\bibinfo{person}{Yuhan Zhao}, \bibinfo{person}{Rui Chen}, \bibinfo{person}{Qilong Han}, \bibinfo{person}{Hongtao Song}, {and} \bibinfo{person}{Li Chen}.} \bibinfo{year}{2024}\natexlab{a}.
\newblock \showarticletitle{Unlocking the Hidden Treasures: Enhancing Recommendations with Unlabeled Data}. In \bibinfo{booktitle}{\emph{RecSys}}. \bibinfo{pages}{247--256}.
\newblock


\bibitem[Zhu et~al\mbox{.}(2024)]%
        {zhu2024reliable}
\bibfield{author}{\bibinfo{person}{Lixi Zhu}, \bibinfo{person}{Xiaowen Huang}, {and} \bibinfo{person}{Jitao Sang}.} \bibinfo{year}{2024}\natexlab{}.
\newblock \showarticletitle{How Reliable Is Your Simulator? Analysis on the Limitations of Current LLM-Based User Simulators for Conversational Recommendation}. In \bibinfo{booktitle}{\emph{WWW}}. \bibinfo{pages}{1726--1732}.
\newblock


\bibitem[Ziarani and Ravanmehr(2021)]%
        {ziarani2021serendipity}
\bibfield{author}{\bibinfo{person}{Reza~Jafari Ziarani} {and} \bibinfo{person}{Reza Ravanmehr}.} \bibinfo{year}{2021}\natexlab{}.
\newblock \showarticletitle{Serendipity in Recommender Systems: A Systematic Literature Review}.
\newblock \bibinfo{journal}{\emph{Journal of Computer Science and Technology}}  \bibinfo{volume}{36} (\bibinfo{year}{2021}), \bibinfo{pages}{375--396}.
\newblock


\end{thebibliography}

\end{document}